\documentclass[iop]{emulateapj}

\usepackage{amsmath}
\usepackage{amsfonts}
\usepackage{bm}
\usepackage{color}
\usepackage{graphicx}
\definecolor{darkgreen}{cmyk}{0.85,0.2,1.00,0.2}

\usepackage{amsbsy}

\input epsf
\usepackage{amsmath,amssymb,subfigure}
\usepackage{graphicx}
\usepackage{epsfig}
\usepackage{color}
\usepackage{ulem}
\usepackage{epstopdf}
\newcommand{\be}{\begin{eqnarray}}

\newcommand{\ee}{\end{eqnarray}}

\newcommand{\jcap}{{\it JCAP}}

\newcommand{\ls}{\mathrel{\raise0.27ex\hbox{$<$}\kern-0.70em \lower0.71ex\hbox{{$\scriptstyle \sim$}}}}

\begin{document}

\title{Searching for Oscillations in the Primordial Power Spectrum: Perturbative Approach
(Paper I)}

\author{P.~Daniel Meerburg$^{1}$,  David N. Spergel$^{1}$ $\&$ Benjamin D.~Wandelt$^{2,3}$}
\affiliation{$^1$Department of Astrophysical Sciences, Princeton University, Princeton, NJ 08540 USA. }
\affiliation{$^2$CNRS-UPMC Univ. Paris 06, UMR7095, Institut d`Astrophysique de Paris, 98bis Bd. Arago, F-75014, Paris, France}
\affiliation{$^3$Departments of Physics and Astronomy, University of Illinois at Urbana-Champaign, Urbana, IL 61801, USA}
\email{meerburg@princeton.edu}
\email{dns@astro.princeton.edu}
\date{\today}

\begin{abstract} 
In this first of two papers, we present a new method for searching for oscillatory features in the primordial power spectrum. A wide variety of models 
predict these features in one of two different   flavors:  logarithmically spaced oscillations and  linearly spaced oscillations. The proposed method treats the oscillations as perturbations on top of the scale-invariant power spectrum, allowing us to vary all cosmological parameters. This perturbative approach reduces the computational requirements for the search as the transfer functions and their derivatives can be precomputed. We show that the most significant degeneracy in the analysis is between the distance to last scattering and the overall amplitude at low frequencies. For models with logarithmic oscillations, this degeneracy leads to an uncertainty in the phase.  For linear spaced oscillations, it affects the frequency of the oscillations. In this first of two papers, we test our code on simulated Planck-like data, and show we are able to recover fiducial input oscillations with an amplitude of a few times $\mathcal{O}(10^{-2})$. We apply the code to WMAP9-year data and confirm the existence of two intriguing resonant frequencies for log spaced oscillations. For linear spaced oscillations we find a single resonance peak. We use numerical simulations to assess the significance of these features and conclude that the data do not provide compelling evidence for the existence of oscillatory features in the primordial spectrum. \end{abstract}

\maketitle
\section{Introduction}

Understanding the physics of the early Universe is  one of the most exciting intellectual challenges of the 21st century. Inflation \citep{1980PhLB...91...99S,1981ZhPmR..33..549M, PhysRevD.23.347,1982PhLB..108..389L} is currently the  most widely studied model of early universe physics. In this model, an as of yet unknown degree (or degrees) of freedom source the exponential expansion of early Universe, redshifting away initial features, including deviations from flatness and pre-inflationary inhomogeneities. Typically, this degree of freedom is a light scalar field which potential energy dominates over all other available degrees of freedom. While the functionally most simple model \citep{2013arXiv1303.5082P,2013arXiv1303.5084P},  a quadratic self interaction, remains within observational bounds \citep{2013arXiv1303.5082P} and is favored by  Occams' razor and entropic reasoning, fundamental theories are unlikely to have such a simple low energy limit.  For example, string theory, the most plausible proposal for UV completion,  has difficulties realizing a single field slow-roll model of inflation (see e.g.~ \cite{2008GReGr..40..565M} for an overview).

Features in the power spectrum are a potential signature of the underlying symmetries that generate inflation. One of these symmetries could be a shift symmetry \citep{2012JCAP...12..036B}, in which the inflaton, composed of pseudo scalar (the axion), obeys a shift symmetry that keeps the action invariant under a discrete symmetry. Inflation itself is realized through small quantum correction to the potential \citep{PhysRevLett.65.3233,2008PhRvD..78j6003S}. Such models lead to oscillations in the primordial power spectrum \citep{2010JCAP...06..009F,2011JCAP...01..017F}. Although we consider these models to be most realistic, other possibilities exist to generate resonance in the primordial spectra. For example, it has been argued that a resonance between negative and positive frequency modes in a pure state Bogolyubov rotation can lead to resonance, both in log space (NPH) \citep{2005tsra.conf....1G} and in linear space (BEFT) \citep{2009JCAP...05..018M}. Recently, a new UV complete model  referred to as unwinding inflation has been proposed \citep{2013JCAP...03..004D}. In this model, log-spaced oscillations are naturally produced when the flux associated with the inflaton scalar unwinds on cycles in compact directions. In two-field models, a bend in field space can also cause oscillations, or features (see e.g  \cite{2011JCAP...01..030A} and more recently \cite{2013JCAP...05..006B}). 

In this paper, we introduce a new method to search for resonance in the CMB power spectrum, with the aim to apply this approach to the recently released Planck data\footnote{http://irsa.ipac.caltech.edu/data/Planck/release} in a companion paper. Similar analysis has been performed in e.g.~\cite{2004PhRvD..69h3515M, 2007PhRvD..76b3503H, 2010JCAP...04..010H, 2010JCAP...06..009F,2011PhRvD..84f3515D, 2012MNRAS.421..369M,2011PhRvD..84f3509B,2013PhRvD..87h3526A,2013JCAP...07..031H} and more recently by \cite{2013arXiv1303.2616P} and the Planck collaboration \citep{2013arXiv1303.5082P}. For completeness, we will consider log spaced oscillations as well as linear spaced oscillations. We will not be concerned with a specific model, although our set-up should allow to put constraints on a variety of models using the results presented here. Our main purpose in this paper is to test our method on simulations and on WMAP 9-year \citep{2012arXiv1212.5225B}  data. For logarithmic spaced oscillations this allows us to compare our findings with previous results and check for consistency.  

The models that we will consider in this paper have the following parametric form:
\be
_1\Delta^2_{\mathcal{R}}(k)  = A_1 \left(\frac{k}{k_*} \right)^{m}\left(1+A_2 \cos [\omega_1 \log k/k_* +\phi_1]\right)\\
\label{eq:powerspectra1}
_2\Delta^2_{\mathcal{R}}(k) = B_1\left(\frac{k}{k_*} \right)^{m}\left(1+B_2 k^n \cos [\omega_2 k +\phi_2]\right)
\label{eq:powerspectra2}
\ee
For example, in axion-monodromy inflation one finds $A_1 = H^2/(8\pi^2\epsilon)$, $m=n_s-1$, $A_2 = \delta n_s$, $\omega_1 = -(\phi_*)^{-1}$ and $\phi_1 = \phi_*$, while for models that compute the effects from a possible boundary on effective field theory (BEFT) predict $B_1 = H^2/(8\pi^2\epsilon)$, $m=n_s-1$, $B_2  = \beta/a_0M$, $n=1$, $\omega_2= 2/a_0 H$ and $\phi_2=\pi/2$. Both initial state modifications and multiverse models \citep{2013JCAP...03..004D} can also produce logarithmic oscillations, while sharp features \cite{2007JCAP...06..023C} result in a power spectrum generate linear oscillations (although the amplitude is typically damped as a function of scale).

This paper is organized as follows. We will discuss some of the complications present in the search for oscillatory features in \S\ref{introduction}. In \S\ref{Method}, we explain how a perturbative approach can improve the search for oscillations, specifically at high frequencies (where high multipole sampling and momentum sampling become more important). We discuss sources of error associated with our approach. We simulate fiducial Planck-like data with and without oscillations and apply our code to this data in \S\ref{simulations} to test the robustness of our code. As a test, we apply our code to the WMAP9  data in \S\ref{WMAP9analysis} for log-spaced oscillations. We discuss our findings and improvement of fit in \S\ref{discussion} and we conclude in \S\ref{conclusion}. 

\section{The search for resonances: the challenge of exploring a highly structured likelihood surface} \label{introduction}
   
Observations of the cosmic microwave background (CMB) provide our best constraint on initial conditions, and provide powerful constraints on $\Lambda$CDM parameters. The CMB power spectrum is  not only sensitive to all 6 parameters ($\Omega_b h^2$, $\Omega_{cdm}h^2$, $\tau$, $A_s$, $n_s$ and $H_0$) and possible extensions to the plain vanilla model \citep{2013arXiv1303.5076P, 2013arXiv1303.5082P}, but also to features in the CMB spectrum.

In most analyses of CMB data, the likelihood surface is well behaved with a shape close to a multidimensional Gaussian.  In this limit, a Monte-Carlo Markov Chain can rapidly explore the likelihood space.  This is not true for models with oscillatory features in the spectrum.  The 
 additional of three new parameters, the amplitude of the correction, the frequency of the oscillation and a  phase, generates a likelihood
surface that is no longer smooth as oscillations can ``line up" with features in the
data produced by either cosmic variance, by noise, or by underlying physics. There are often many isolated minima, particularly when the the frequency is high and the amplitude small. 
While Markov Chains will converge in the limit of very many steps, in practice this can take a very long time.

There are several possible approaches to searching a complex likelihood surface:
\begin{itemize}
\item  We could try to sample of dense grid of  possible parameter values.  For a full fledged grid search, the number of samples grows as $N_1\times N_2 .... N_k$ with $N_i$ samples for $k$ parameters. Suppose we want to compute a ten points for each parameter (which is really low), with our 9 parameter model we would end up with $10^9$ points. Computing a single power spectrum up to $l=2500$ typically costs a few second on a single CPU. Therefore we find that this computation would take us over 300 years of CPU time!  
\item A more promising approach is to use  more advanced MCMC routines such as Multinest \citep{2009MNRAS.398.1601F}. This technique has been recently applied to this problem, although with most parameters set to their best-fit values \citep{2013arXiv1303.2616P, 2013arXiv1303.5082P}. With only 3 free parameters, multi-nest is not much faster than  a grid search.
\item A reasonable compromise is to grid sample only the parameters that require a close inspection of the likelihood (e.g. the frequency, amplitude and phase) while keeping all other parameters fixed close to their best-fit values based on the MCMC without oscillations. This approach has been attempted by \cite{2010JCAP...06..009F} and \cite{Meerburg:2011gd}. In these examples, one typically finds several frequencies that can lead to an improved fit with $\Delta \chi_{\mathrm{eff}}^2 \sim \mathcal{O}(10)$. After the grid search, one can apply an MCMC keeping the best-fit frequency fixed, while varying the remaining parameters, including the phase and the amplitude of the oscillatory correction. In the ideal scenario, where the grid parameters are only marginally correlated with the MCMC parameters, this approach should be reasonably accurate.
\end{itemize}

In our analysis, we pursue an alternative, hybrid approach. We note that the likelihood surface
at fixed frequency is smooth and does not have large numbers of multiple minima.  Thus, by running chains in the eight dimensional space at fixed
frequency,
we avoid many of the pitfalls of trying to explore the nine dimensional space.  While this approach does require that we run chains at each
frequency, the next subsection outlines our approach for speeding the computation of the angular power spectrum for rapidly oscillating power spectra.

\section{Perturbative approach} \label{Method}

In this subsection, we introduce a perturbative approach for rapidly evaluating the angular power spectrum.  

The predicted angular power spectrum, $C_l$, is an integral over the primordial fluctuations weighted by a transfer function,  $\Delta_l^T(k)$,
\be
C_l = \frac{2 }{\pi}\int_0^{\infty} \frac{d k}{k}  \Delta^2_{\mathcal{R}}(k)(\Delta_l^T(k))^2
\label{eq:cl}
\ee  
Evaluating the transfer function is the most time consuming part of the calculation. When the power spectrum is smooth, we can compute the transfer function for a coarse grid in $\ell$ and and integrate over sufficient resolution in $k$.  However, when there are a large number of primordial oscillations in $\Delta^2_{\mathcal{R}}(k)$, there are a large number of oscillations in $\mathcal{C}_{\ell}$, hence one needs a high $\ell$ resolution (every time we change the parameter values that determine the geometry of the Universe). For log space oscillations this computational burden can partly be mitigated by sampling $\ell$ space adaptively. For linear space oscillations and for rapid log-spaced oscillations this is no longer true, and for an accurate $C_{\ell}$ one needs to compute the transfer function for all $\ell$ up to $\ell_{\mathrm{max}}$.  

Since the perturbations in the power spectrum are small and 
and since the transfer function does not depend on initial conditions but
only on the properties of the $z \sim 1100$ universe (the baryon density and the matter density) and effects along the line of sight (the distance to
the surface of last scatter and the optical depth), we  can accurately compute the angular power spectrum by treating the oscillatory term as
small and expanding the transfer function in a Taylor series.

\begin{figure}[htbp] 
   \centering
   \includegraphics[width=3in]{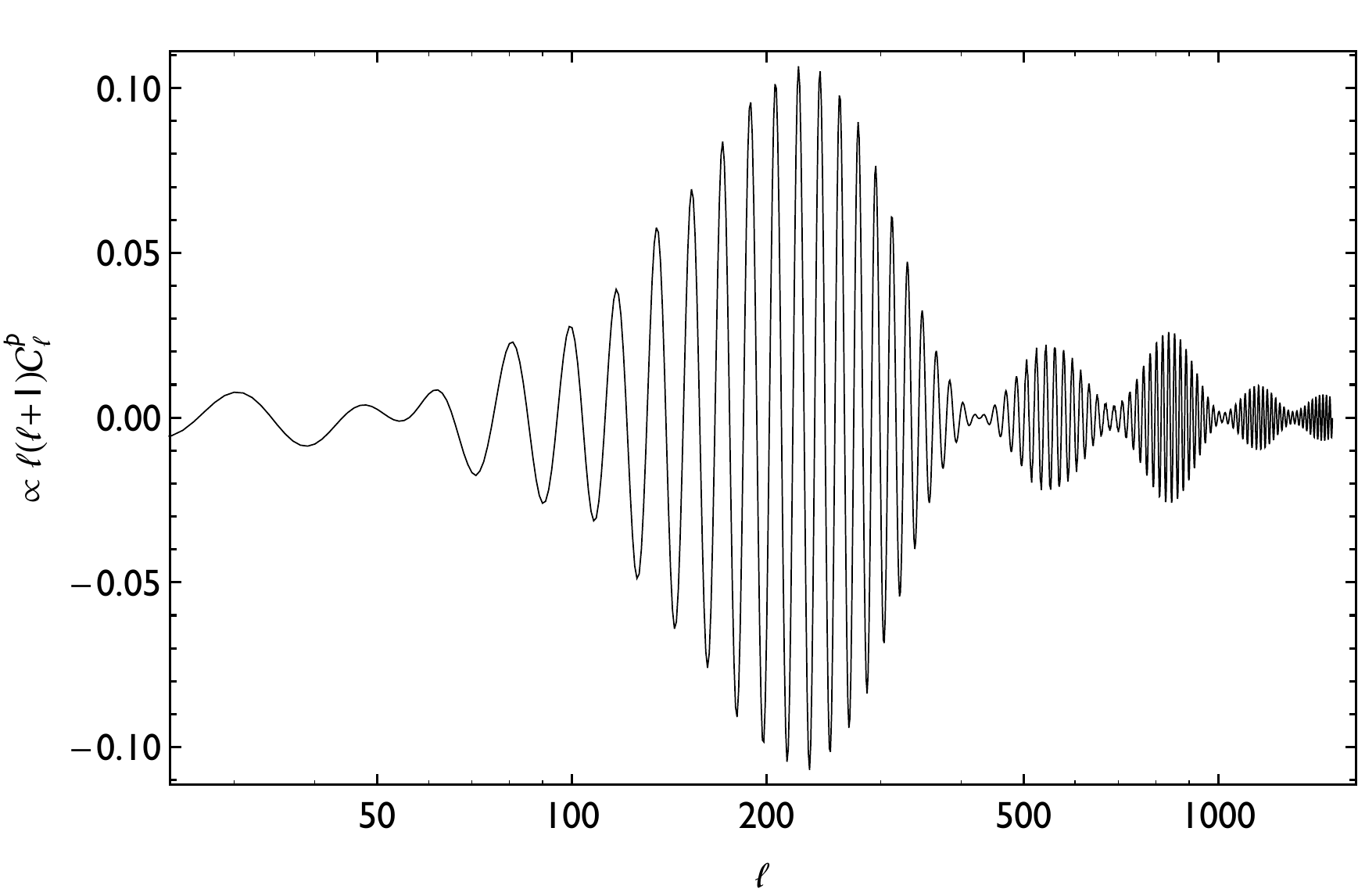} 
   \caption{Example of the perturbed power spectrum of linearly spaced oscillations with $\omega_2 = 5 \times10^3$. One can roughly estimate the wavelength through $\lambda_{\ell} = 2\pi \Delta \eta/\omega_2$, with $\Delta \eta$ the conformal distance to last scattering. For this example we therefore find $\lambda_{\ell} \simeq18$. Since we expect we could at best resolve $\lambda_{\ell}=2$, this puts an upper limit to $\omega_2\leq 40000$. Note that the normalization is arbitrary.}
   \label{fig:cposcillations}
\end{figure}

\begin{figure}[htbp] 
   \centering
   \includegraphics[width=3in]{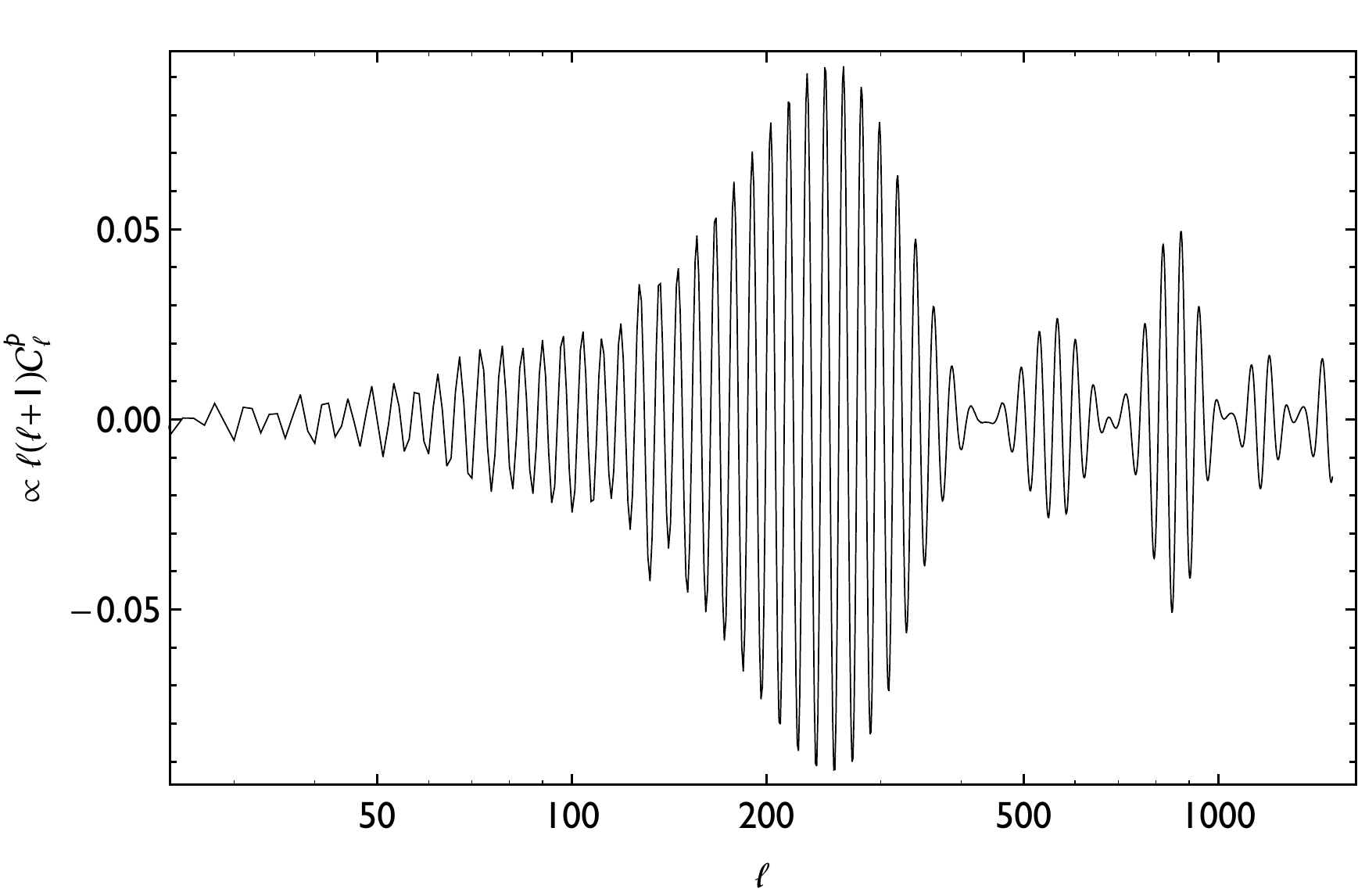} 
   \caption{Example of the perturbed power spectrum of logarithmically spaced oscillations with $\omega_1 =100$ (used in simulations). The number of oscillations per $\Delta \ell$ interval increases from low to high $\ell$. Therefore the observabilty of these modulations will depend on $\ell_{\mathrm{max}}$. A frequency $\omega_1 = 100$ roughly corresponds to a wavelength of $\lambda_{\ell} = 14$ at $\ell =200$. }
   \label{fig:cplogoscillations}
\end{figure}

Let us consider the following model for illustration
\be
_1\Delta^2_{\mathcal{R}}(k)  &=& A_1 \left(\frac{k}{k_*} \right)^{m}\left(1+A_2 \cos [\omega_1 \log k/k_* +\phi_1]\right)\nonumber \\
&= & A_1 \left(\frac{k}{k_*} \right)^{m} + \alpha \left(\frac{k}{k_*} \right)^{m}  \cos [\omega_1 \log k/k_*]+\nonumber \\
&& \beta \left(\frac{k}{k_*} \right)^{m}  \sin [\omega_1 \log k/k_*]
\ee
Here we explicitly decided to expand the phase into two oscillating components, with $\alpha = A_1 A_2\  \cos \phi_1$ and $\beta = -A_1 A_2   \sin \phi_1$ (this allows us to vary this parameter after precomputing the integral of Eq.~\eqref{eq:cl}). We know from observations and from theoretical bounds that the oscillations can never exceed the non-oscillating part. Setting $m=n_s-1 \simeq 0$ the total $C_l$ can be written as
\be
C_{\ell} &\equiv & C_{\ell}^u+ C_{\ell}^p\nonumber\\
&=&\frac{2}{\pi}\int_0^{\infty} \frac{d k}{k} \left[A_1+ \alpha \cos [\omega_1 \log k/k_*]+\right.\nonumber\\
&& \left.\beta \sin [\omega_1 \log k/k_*]\right](\Delta_l^T(k))^2
\ee
Because the correction to the unperturbed spectrum, $C^{u}_l$, is small, we can assume that any estimates to the actual value of the late time parameters will be relatively insensitive to the `enveloped' shape (as shown in Fig. \ref{fig:cposcillations} and \ref{fig:cplogoscillations}) of the oscillatory part.  We can Taylor expand in that parameter around the best-fit value in the unmodulated power spectrum, i.e. 
\be
(\Delta_l^T(k))^2 &=& (\bar{\Delta}_l^T)^2+2 \bar{\Delta}_l^T\sum (\Theta_i - \bar{\Theta}) \bar {\Delta}_{l,\Theta_i}^T+\mathcal{O}((\Theta_i-\bar{\Theta})^2)
\ee
where $\bar{\Theta}$ is the best-fit value of the $\Theta_i$ parameter for an unmodified power spectrum, $\bar{\Delta}_l^T$ is the transfer function computed with $\Theta = \bar{\Theta}$, and  $\bar {\Delta}_{l,\Theta_i}^T$ represents the  derivative of the transfer function w.r.t. to the parameter $\Theta_i$, evaluated at $\Theta = \bar{\Theta}$. We consider these corrections second order, since they multiply the amplitude of the perturbed part, with first order corrections to the transfer function. As explained, the best-fit parameters $\bar{\Theta}$ can be obtained relatively fast with a single cosmomc \citep{2002PhRvD..66j3511L}  run. 
The expansion allows us to precompute the transfer functions. Once these values have been determined (for a given data set(s)) we can now precompute the corrections, for a large number of frequencies ($\omega_1$), i.e. 
\be
C_{\ell}^p & = &\frac{\pi}{2}\int_0^{\infty} \frac{d k}{k}\left[\alpha \cos [\omega_1 \log k/k_*] \right. \nonumber \\
&& \left. +\beta \sin [\omega_1 \log k/k_*]\right](\Delta_l^T(k))^2 \nonumber \\
&= &\alpha \frac{\pi}{2} \left[\int_0^{\infty} \frac{d k}{k} \cos [\omega_1 \log k/k_*](\bar{\Delta}_l^T)^2 +\right. \nonumber\\
&& \left.  2\sum (\Theta_i - \bar{\Theta}_i)\int_0^{\infty} \frac{d k}{k} \cos [\omega_1 \log k/k_*]\bar{\Delta}_l^T \bar {\Delta}_{l,\Theta_i}^T \nonumber \right] +\nonumber\\
&& \beta \times ... + \mathcal{O}((\alpha+\beta)(\Theta_i-\bar{\Theta})^2)\nonumber \\
&\approx&\bar{C}_{\ell}^{p(\alpha)} +\bar{C}_{\ell}^{p(\beta)}+ \sum (\Theta_i-\bar{\Theta}_i)( \bar{C}_{\ell,\Theta_i}^{p(\alpha)}+ \bar{C}_{\ell,\Theta_i}^{p(\beta)}) 
\label{eq:expansion}
\ee
In the last line we used the commutation of the derivative operator and the integration for continuous functions.
We will argue that for our purposes we can truncate this expansion at zeroth order in  $(\Theta_i-\bar{\Theta}_i)$ for all $i$.
The last line is {\it general}, in the sense that it should hold for any oscillatory correction, as long as we assume the amplitude is small. We precompute the integrals in the equation above (for each $\ell$ up to some $\ell_{max}$ related to the angular resolution of the experiment) and sufficient $k$ with fixed $\Lambda$CDM parameters for a large set of $\omega_1$ (derived from the best-fit without oscillations). Even for high frequencies, we can parallelize our code and compute 3000 spectra in less than 12 hours on a single node with 12 cores. For any given data set, we only have to do this once, and in principle there are two types, related the form of the two example power spectra in Eqs.~\eqref{eq:powerspectra1} and ~\eqref{eq:powerspectra2}. If we want to include higher order corrections, we can compute the derivatives $\bar{C}_{\ell,\Theta_i}^{p}$ (we leave $\alpha$ and $\beta$ as free parameters). Again these derivatives evaluated at the best-fit point can be precomputed at a cost of very little additional CPU time. 


\subsection{Sources of error in the approximation}
 
There are two distinct sources of error in the approximation Eq.~\eqref{eq:expansion}. The first source of error is caused by expanding about the wrong {\it model parameters}, while the second source of error is caused by truncating the series at too low an order. Although  these two sources are not completely independent, for reasons of clarity, we will discuss them separately. In principle, both sources of error can be reduced by considering higher order terms in the expansion. We would like to stress that our approach is generally more accurate than most attempts in the literature since in most cases  all cosmological parameters are held fixed to their best-fit values. 

The first error is a consequence of fixing the cosmological parameters to the best-fit values derived from a fit {\it without} oscillations. Ignoring the fact that the best fit may change in the larger model that includes the oscillations this approximation can introduce an error in the derived oscillatory parameters as we will show below.

If the model spectrum (i.e the oscillating spectra of Eq.~ \eqref{eq:powerspectra1}) is the true spectrum, our approximation results in an error in  the calculation of $C_{\ell}$ that is proportional to the derivatives of the perturbed part with respect to the parameters of interest. This is the second source of error and can lead to errors in all derived parameters. Interestingly, the presence of oscillations could {\it improve} the measurement of certain parameters, because of a denser sampling of the transfer functions. We will show that this effect could in principle lead to a larger truncation error in these parameters, but for small values of the primordial amplitude they should stay within the 2$\sigma$ bound of the parameter constraint without oscillations. Therefore this error is relevant only when there exists compelling evidence for an oscillation. Extending the expansion to higher order in these parameters can reduce this error.  

\begin{figure}[htbp] 
   \centering
   \includegraphics[width=3in]{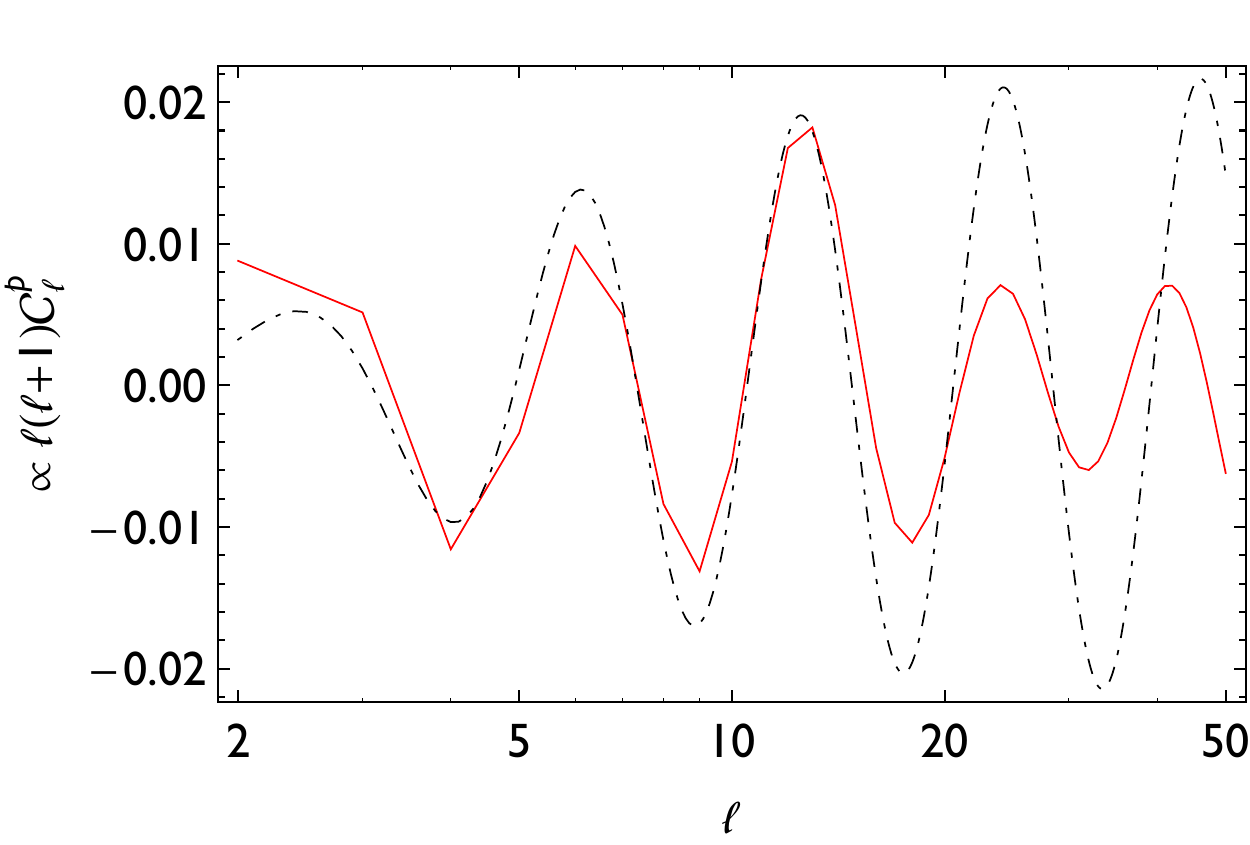} 
   \caption{Comparison between analytical approximation (dot dashed) at low multipole $\ell$ and exact numerical solution (solid, red). The analytical approximation traces the the numerical result closely at the lowest $\ell$ but quickly starts  deviate at high $\l$. The same is true for higher frequency oscillations. }
   \label{fig:anavstrue}
\end{figure}

Let us consider the following example, to clarify the first of the two sources of error. 
We can use the low $\ell$ approximation of the transfer functions to derive an analytical result for log spaced oscillations, i.e. the monopole solution without integrated Sachs-Wolfe effect is projected through
\be
\Delta^T(k) \simeq \frac{1}{3} j_{\ell}(k \Delta \eta),
\label{eq:SWtransfer}
\ee
with $\Delta \eta$ the (conformal) distance to last scattering and $j_{\ell}(k)$ the spherical Bessel functions. The perturbed $C_{\ell}^p$ can therefore be approximated with \citep{2004PhRvD..69h3515M}
\be
C_{\ell}^p \simeq \frac{2}{9\pi} \int dk \cos [\omega_1 \log k + \phi] j_{\ell}^2(k \Delta \eta).
\label{eq:cllana}
\ee
We have absorbed the $1/k_*$ in the phase $\phi$ which we  set to zero for convenience (it is straightforward to put it back in). This integral can be performed analytically and we find 
\be
C_{\ell}^p \simeq \frac{1}{36 \sqrt{\pi }}  \left[\frac{ (\Delta \eta)^{ i \omega_1 } \Gamma \left(\frac{i \omega_1 }{2}+1\right) \Gamma \left(\ell-\frac{i \omega_1
   }{2}\right)}{\Gamma \left(\frac{i \omega_1 }{2}+\frac{3}{2}\right) \Gamma \left(\ell+\frac{i \omega_1 }{2}+2\right)}+\mathrm{c.c}\right].\nonumber\\
   \label{eq:cllowl}
\ee
This solution is plotted against the exact solution in Fig.~\ref{fig:anavstrue}.

Looking at Eq.~\ref{eq:cllana} the comoving wavelength in the argument in the transfer function explicitly depends on the distance to the last scattering surface. This can be reabsorbed into the integral via a transformation $\tilde{k}\rightarrow k\Delta \eta$. Effectively, for the log spaced oscillations above, we get a phase shift $\Delta \phi \sim -\omega_1\log \Delta \eta$ (appearing as $(\Delta \eta)^{ i \omega_1 } $ in Eq.~\eqref{eq:cllowl}). Although we can not perform the linear spaced analog analytically, a similar stretching of the comoving wavelength results in a reduction of the primordial frequency $\omega \sim \omega/\Delta \eta$. This is physically intuitive as the start of the oscillation (phase) and the effective number of oscillations (the frequency) depend on the line of sight distance. Since 
 \be
 \Delta \eta = \int_{a_*}^1 \frac{da}{a^2H(a)},
 \ee 
this distance depends on late time $\Lambda$CDM parameters alone. Consequently, when applying the approximation we use to analyze the data, by fixing the late time cosmological parameters in the precomputed perturbed spectra to their best-fit values, adding oscillations can lead to a deviation between the actual distance to last scattering and the precomputed one. We have confirmed this effect through simulations; generating data with an exact spectrum but random noise, results in shifts of the derived parameters of order $\sigma$. When applying precomputed spectra to these generated mock data, where the precomputed spectra are based on the exact values of parameters, we find a shift in the phase for log spaced oscillations and a shift in the frequency for linear spaced oscillations. These shifts are reduced when we render the precomputed spectra using derived parameters instead of exact parameters. Consequently, besides expanding the precomputed part to higher order one could reduce this error through iteration; take the best-fit-value, generate the transfer functions, apply the approximated model to the data, derive the updated values, and recompute the transfer function, apply those to the data, etc. 

Now lets us quantify the second source of error,  the deviation  caused by applying the truncated model to the data. Suppose the true  model is one with oscillations, 
\be
C_{\ell} &\equiv & C_{\ell}^u+ C_{\ell}^p. \nonumber \\
\ee
To zeroth order in the expansion, the bias w.r.t. the actual $C_{\ell}$ is proportional to 
\be
\Delta C_{\ell} \simeq \sum \frac{\partial C^p_{\ell}}{\partial \Theta_i} \Delta \Theta_i.
\ee
The bias drives parameters away from the actual values. The validity of the expansion will be determined to what extent the perturbed part actually contributes to the total $\chi^2$. If there are no oscillations the bias disappears. For that reason, we set the phase to 0, and define 
\be
 C^p_{\ell} = A \tilde{C}^p_{\ell},
\ee
where $A$ now is the (phase absorbed) amplitude of the primordial oscillatory correction. 
The following quantity measures the contribution of the oscillatory correction to the parameter log-likelihood
\be
A^2 \Delta \Theta_i F_{ij} \Delta \Theta_j =\epsilon,
\ee
where 
\be
F_{ij} = \sum (2\ell+1) \frac{1}{\left(C_{\ell}+ N_{\ell}\right)} \frac{\partial \tilde{C}^p_{\ell}}{\partial \Theta_i}\frac{\partial \tilde{C}^p_{\ell}}{\partial \Theta_j}.
\label{eq:fisherbound}
\ee
$N_{\ell}$ is the noise of the experiment (in principle one should use the data covariance). If $\epsilon<1$  it might be necessary to rerun the analysis and include higher order terms (or run a non-perturbative chain). For example Eq.~\eqref{eq:fisherbound} can be determined for chain (parameters) associated with a a non-zero oscillatory amplitude (if the chain prefers a zero amplitude, the bound is satisfied automatically). The parameters in the chain must be compared to the best-fit input parameters used to generate the transfer functions (i.e. the $\Delta \Theta$).  

The Fisher matrix (Eq.~\eqref{eq:fisherbound}) depends on the derivatives of the perturbed part with respect to the $\Lambda CDM$ parameters. Previously we argued that a clear source of error in the derived oscillatory parameters was driven by projection from last scattering, leading to error in the phase (log spaced oscillations) and the frequency (linear spaced oscillations). However, the frequency of log spaced oscillations can also be affected by the transfer functions (albeit less obviously). Any derivatives with respect to the parameters that influence the frequency will therefore increase in amplitude as you increase the frequency. Effectively what is happening is that any presence of oscillations measures the transfer functions that depend (predominantly) on $\Omega_b h^2$, $\Omega_{dm}h^2$ and $H_0$ more accurately. Therefore we expect that as the frequency increases, our accuracy of these parameters should increase, while the accuracy of other parameters will get {\it worse} (i.e. $n_s$, $A_s$ and $\tau$). We would like to emphasize that this error is not relevant for recovering an oscillatory signal (it will not have an effect on the ability to recover the frequency or amplitude of the input spectrum as we will see in the next section), but is relevant if one wants to improve the measurement of other parameters. We will compute this bound for the best-fit chain from WMAP in section \S\ref{WMAP9analysis} Fig.~\ref{fig:fij}. Again, we would like to emphasize that these errors of measure are present when you fix all cosmological parameters as well.

\section{Simulations} \label{simulations}

The purpose of simulations is twofold. They test the robustness of our code (given the possible errors given above) and they evaluate the significance of any measured improvement given the signal.



\subsection{Log-spaced oscillations}

\begin{table} 
\centering 
\begin{tabular}{l c c c c c |} 
\toprule 
& \multicolumn{4}{c}{{\bf Planck-like data}} \\ 
Channel & 143 GHz & 100 GHz & 70GHz &\\ 
FWHM[arcmin] & 7.1 & 10 & 14 &\\ 
$\sigma_T$ [$\mu$K p/p] & 6.0 & 6.8 & 12.8 &\\ 
$\sigma_P$  [$\mu$K p/p] & 49 & 49 & 49 & \\ 
\end{tabular}
\caption{Noise statistics used to generate Planck-like data.} 
\label{tab:pnoise} 
\end{table}

We generate Planck-like data {\it with exact spectra}. The noise statistics are shown in Table \ref{tab:pnoise}, with 3 mock channels and WMAP polarization noise. We slightly modified version of the code provided by \citep{2006JCAP...10..013P} to generate the maps. 
We create fiducial spectra with $A_2=0.1,\;0.05$ and $0.01$. We have performed a high sampling of a fiducial frequency at $\omega_1 = 100$ (see Fig~\ref{fig:cplogoscillations}), with a total of 100 samples. In addition we also performed a low resolution (20 steps in frequency space) sampling with 3 mock spectra at $\omega_1=210$ and at $\omega_1=30$. Maps are generated with the same random seed for each frequency range. 

\begin{figure}[htbp] 
   \centering
   \includegraphics[width=3.55in]{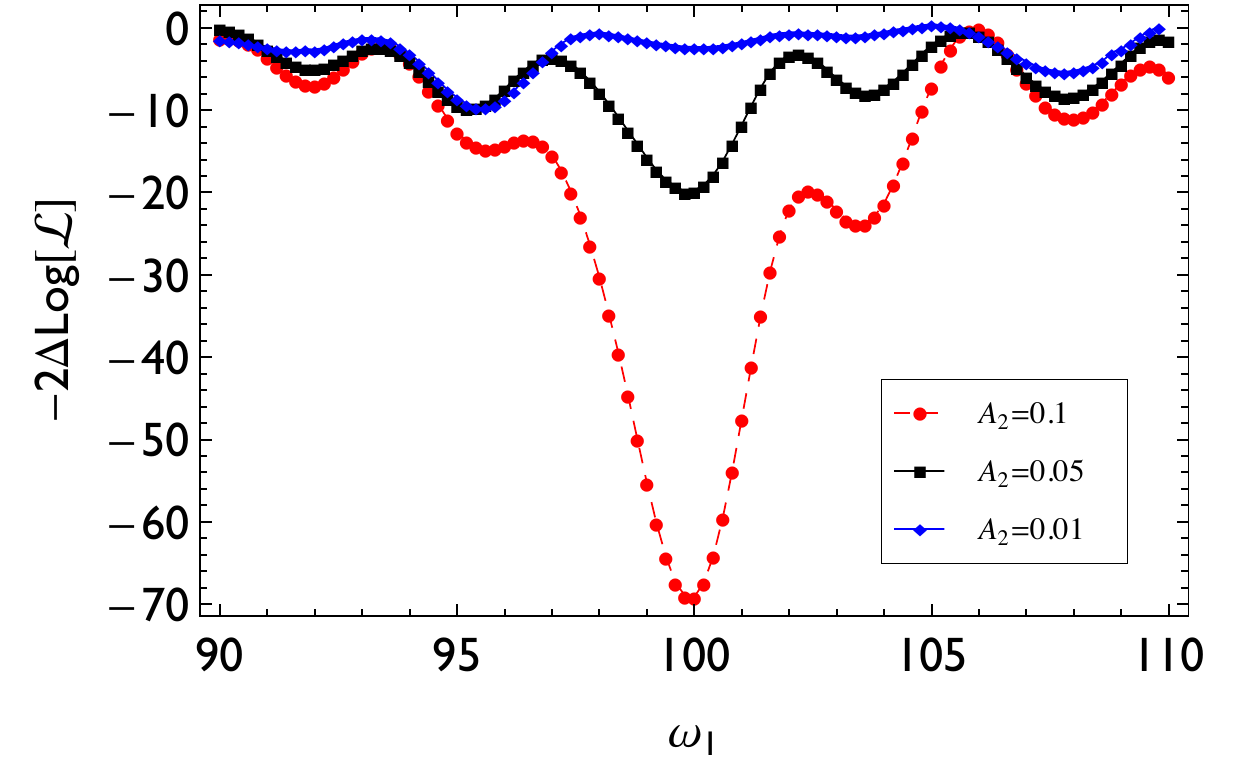} 
   \caption{Improvement of fit versus $\omega_1$ for several input amplitude's. $A_2=0.1$ and $0.05$ are recovered, while $A_2=0.01$ is not. The `oscillations' are a consequence of the noise (which is the same for all 3 simulations). It is clear that features in the noise can amplify and de-amplify some of the signal. }
   \label{fig:simlog100}
\end{figure}

Fig.~\ref{fig:simlog100} shows  $-2 \log \mathcal{L}$ improvement as a function of frequency derived from fiducial maps with $\omega_1=100$, $A_1=.1, 0.5$ and $0.01$ and $\phi_1=0$. We sampled around the fiducial frequency $\omega_1 \pm 10$ to show how the improvement changes as you get further away from the input value. Note foremost that the algorithm recovers the fiducial frequency if $A_2 = 0.05-0.1$. For a amplitude of $A_2 = 0.01$ we find that improvement to the fit is (mostly) due to a fitting of the noise and primordial frequency.  We conclude this on the basis that we neither recover the fiducial amplitude nor the fiducial frequency, and the best-fit improvement doe not coincide with the input spectra. Furthermore we generated mock data with no signal, but the same random noise, and found almost exact overlap with a fiducial map with $A_2=0.01$.  It is also clear that the oscillating pattern of the improvement is consistent for all 3 spectra, which is a consequence of using the same noise seed. It also confirms that the presence of noise can amplify a potential signal.  More importantly, the improvement over a wide range of frequencies is $-2 \Delta \log \mathcal{L} \sim 10$ for $A_2=0.01$, which tells us that it is probably impossible to distinguish between oscillations with an amplitude $A_2\sim0.01$ and the noise using Planck alone. This analysis shows that our approximation, using precomputed transfer functions, works, even though the mock spectra were generated using the exact spectra. We will further comment on these findings in \S\ref{discussion}.

\begin{figure}[htbp] 
   \centering
   \includegraphics[width=3.55in]{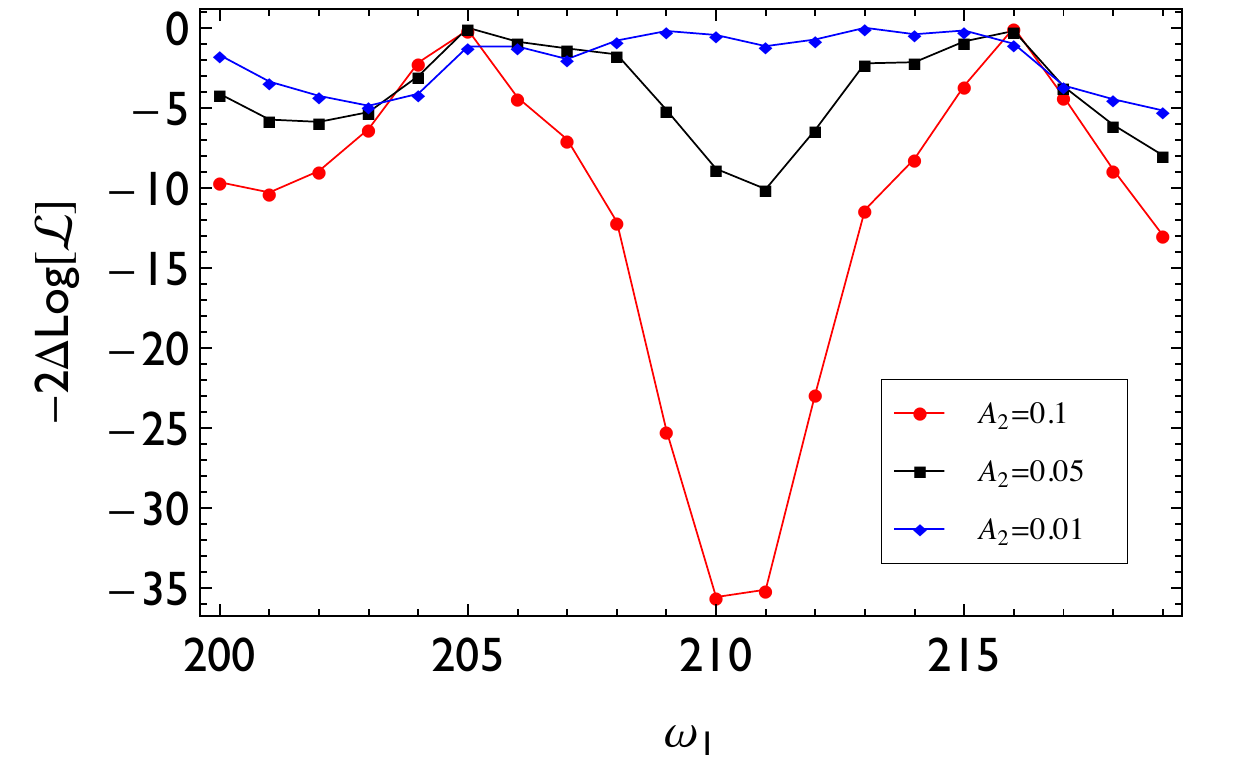} 
   \caption{ Frequency versus the improvement of fit with primordial frequency $\omega_1=210$.}
   \label{fig:simlog210}
\end{figure}

For the high frequency mock data $\omega_1 =210$ (Fig.~\ref{fig:simlog210}), we find that typical improvement in  $-2 \Delta \log \mathcal{L}  $ is smaller, which we attribute to the fact that you lose effective amplitude through projection. Furthermore, in this case there seems to be a small shift in the best-fit frequency related to the input value, although even at low sampling of $\Delta \omega_1 = 1$ we recover a frequency within $1\sigma$ of the input value. 

\begin{figure}[htbp] 
   \centering
   \includegraphics[width=3.55in]{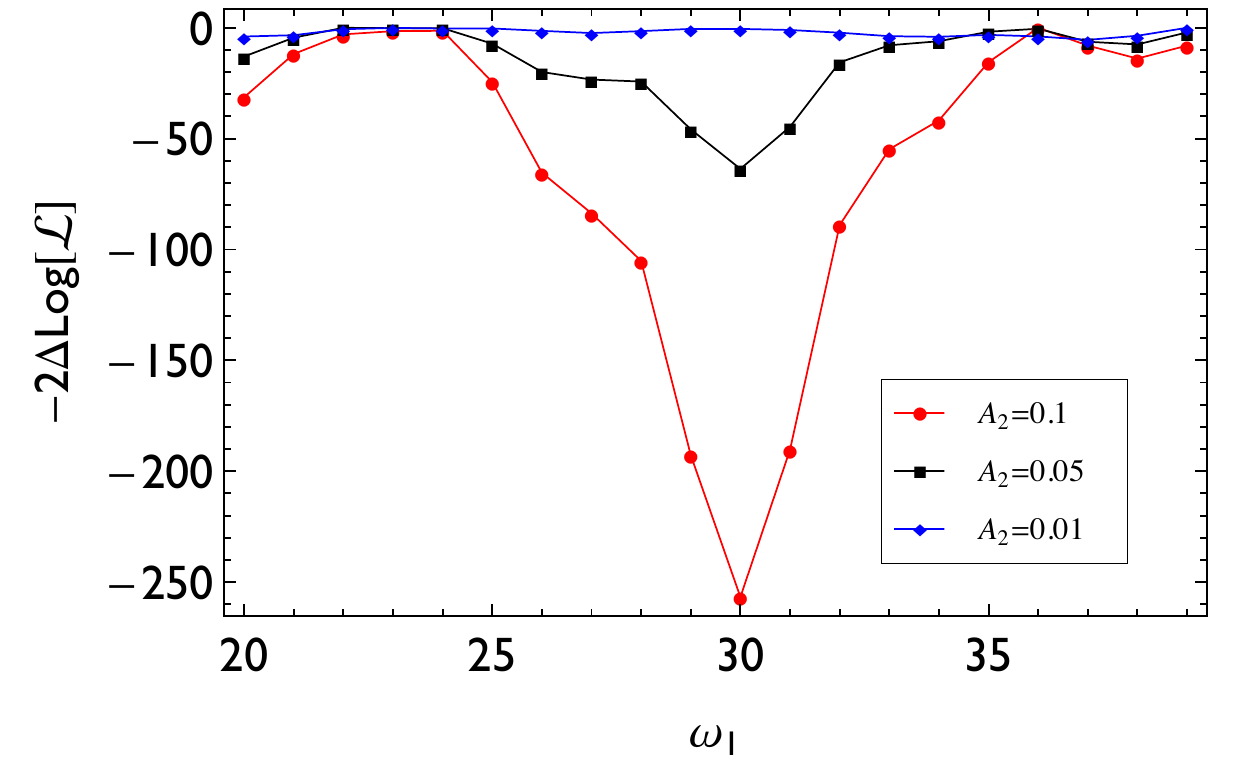} 
   \caption{Frequency versus the improvement of fit with primordial frequency $\omega_1=30$.}
   \label{fig:simlog30}
\end{figure}

For the low frequency mock data $\omega_1 =30$ (Fig.~\ref{fig:simlog30}), we obtain a much bigger improvement in $-2 \Delta \log \mathcal{L}  $. Such a large improvement was expected because projection keeps most of the amplitude of the primordial feature invariant. 

\subsection{Linear-spaced oscillations}
\begin{figure}[htbp] 
   \centering
   \includegraphics[width=3.55in]{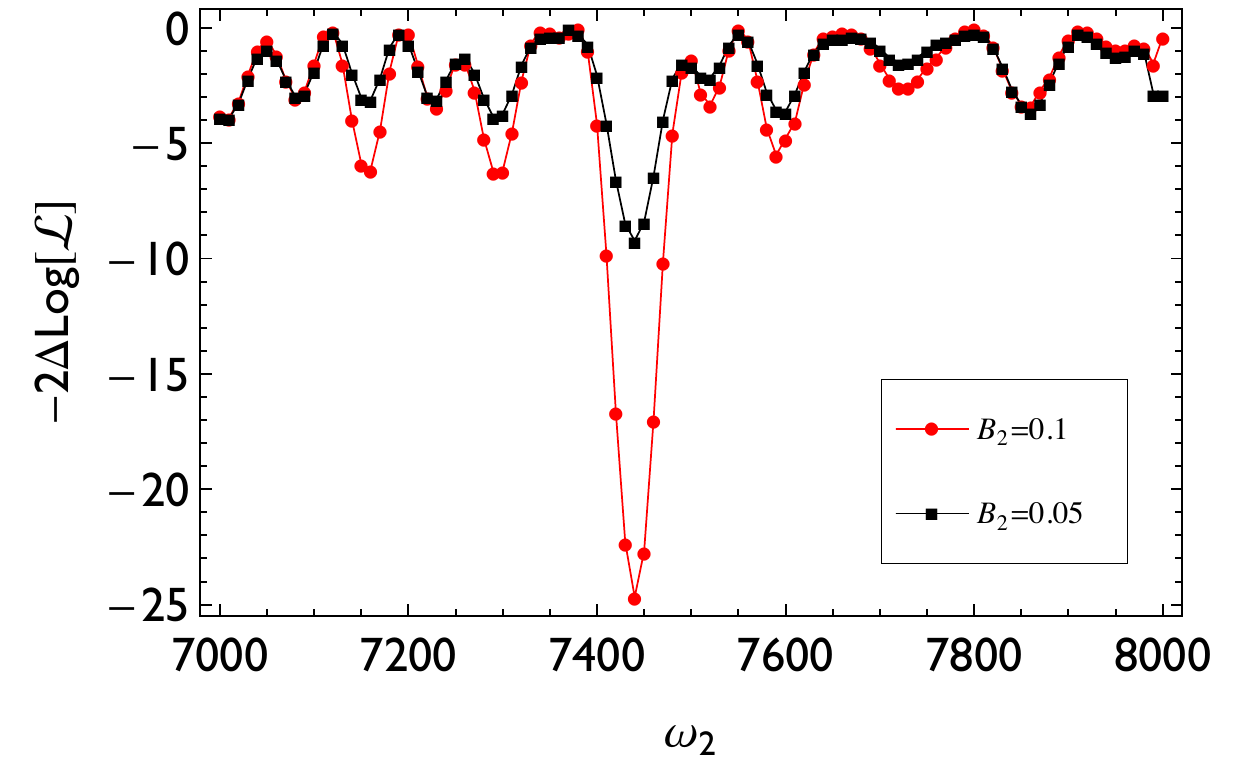} 
   \caption{ Frequency versus the improvement of fit with primordial frequency $\omega_2=7500$. A primordial amplitude below $B_2=0.5$ at these frequencies is very hard to disentangle form the noise. Also note that again the noise is boosted by the presence of the oscillation. }
   \label{fig:simlin7500}
\end{figure}

For linear spaced oscillations we generated 2 maps, with $\omega_2=7500$ and $B_2 = 0.1$ and one with $0.05$ (we have already seen that amplitudes of order $0.01$ are indistinguishable from features in the noise). The result of our blind analysis of these maps is shown in Fig. \ref{fig:simlin7500} where we plotted the improvement of fit versus frequencies. One important  observation is that indeed our recovered frequency has shifted with respect to the input frequency, which was expected. The improvement of the fits is comparable to the high frequency log space simulation, with a best-fit that improves compared to no oscillations with $-2 \Delta \log \mathcal{L} = -25 $. Although the improvement is still large compared to the noise within the search domain for $B_2=0.05$, we will later show that a typical improvement  from the noise is expected to be of the order of $-2 \Delta \log \mathcal{L} \sim -10 $.

\section{WMAP9  Analysis} \label{WMAP9analysis}

\subsection{Log-spaced oscillations}

\begin{figure}[htbp] 
   \centering
   \includegraphics[width=3.4in]{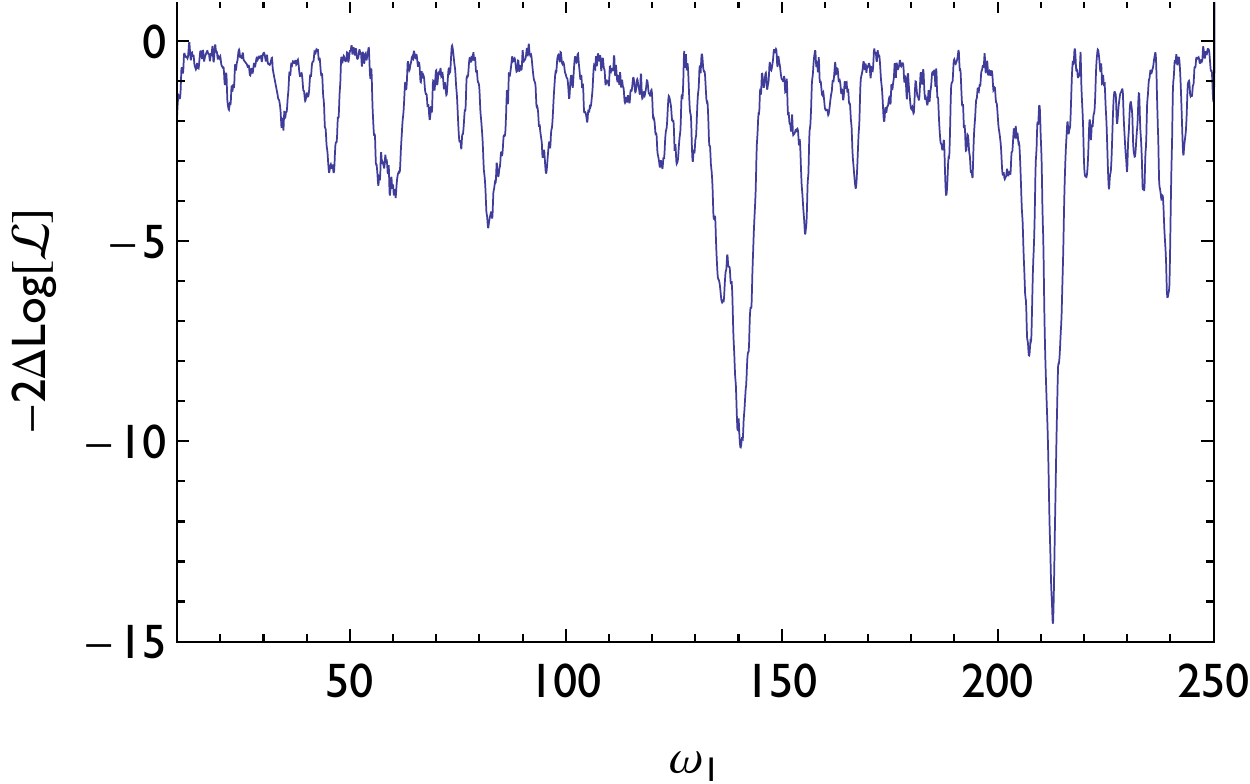} 
   \caption{The improvement of fit for 1201 frequencies in WMAP9  data. Two peaks earlier identified in \citep{2013arXiv1303.2616P} are clearly visible. }
   \label{fig:wmapchi}
\end{figure}

\begin{figure*}
  \centering
  \subfigure{\includegraphics[width=0.45\textwidth]{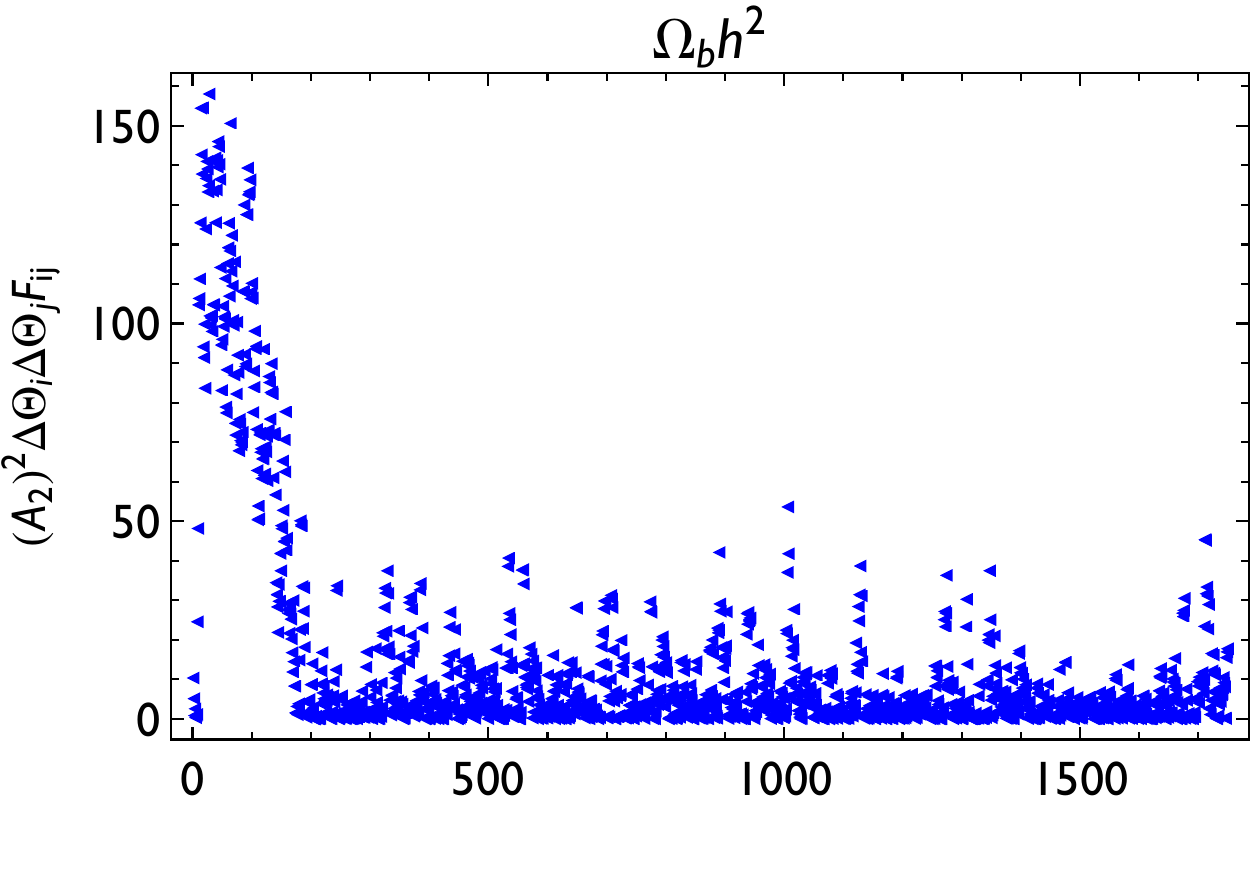}} \qquad
  \subfigure{\includegraphics[width=0.45\textwidth]{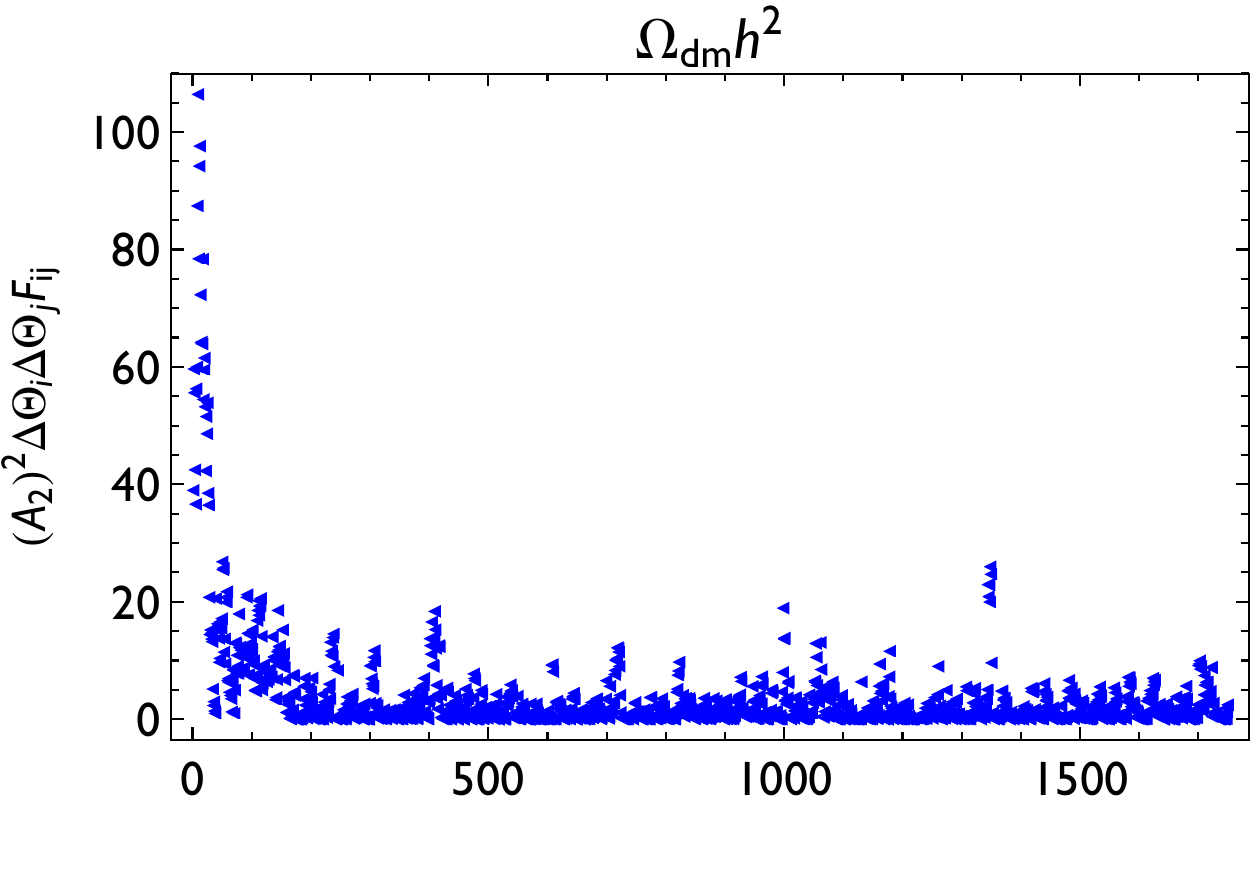}}\\
  \subfigure{\includegraphics[width=0.45\textwidth]{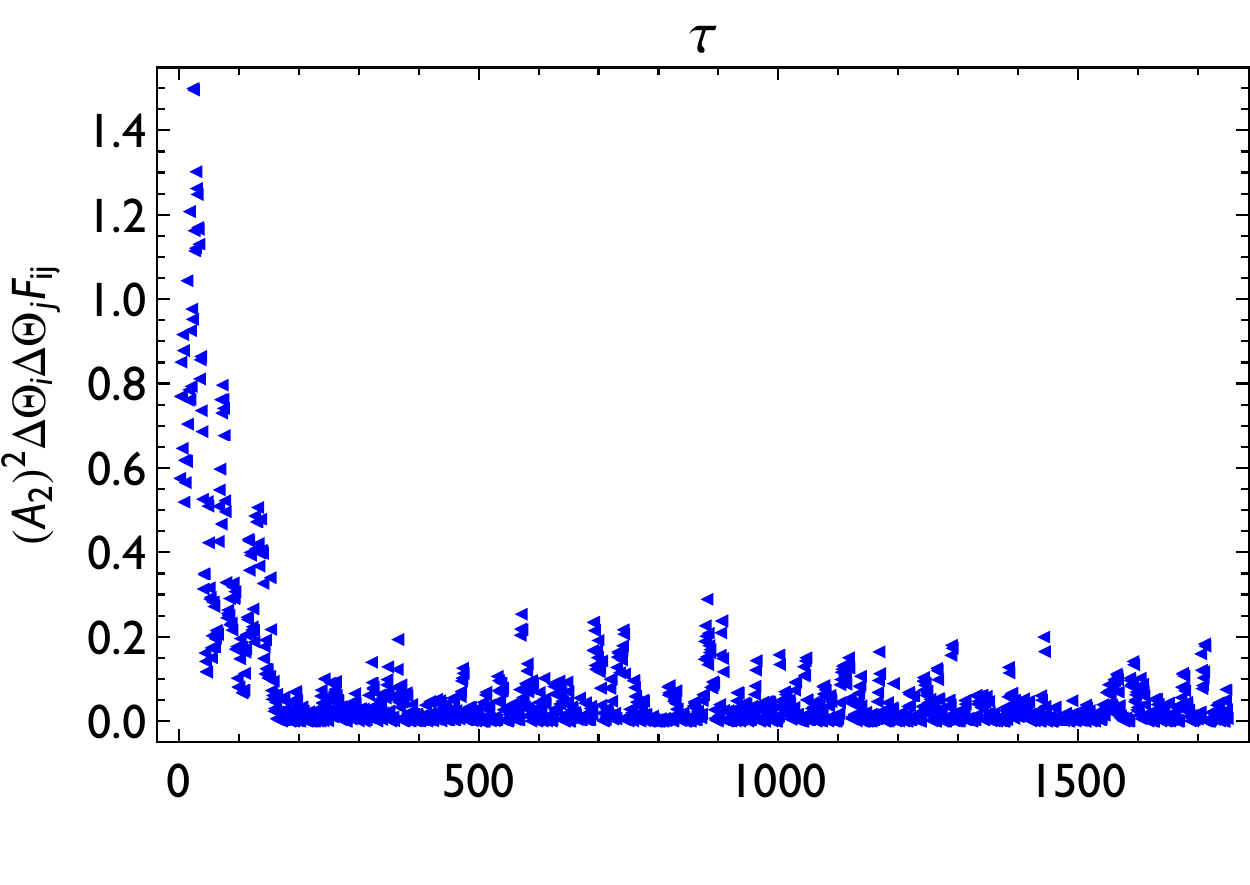}}\qquad
  \subfigure{\includegraphics[width=0.45\textwidth]{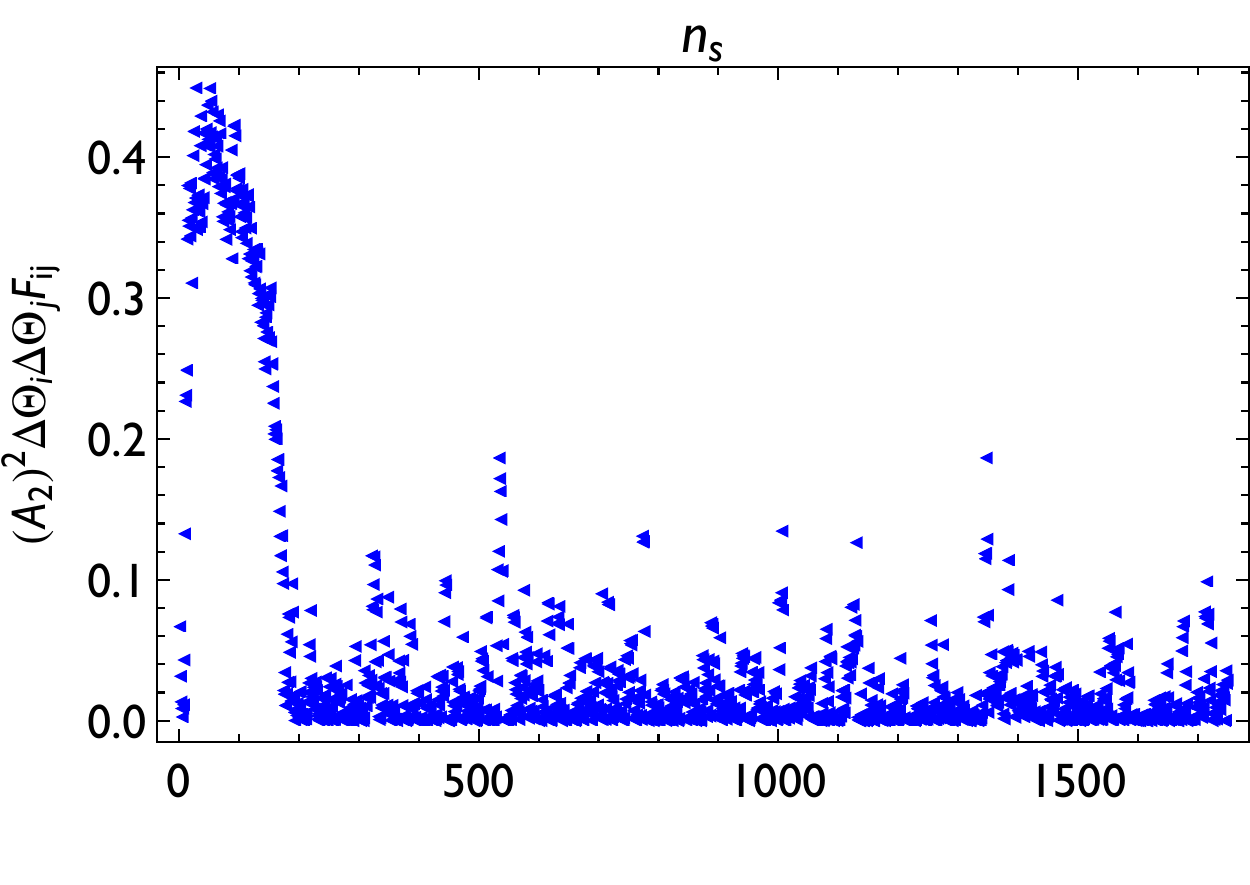}}\\
   \subfigure{\includegraphics[width=0.45\textwidth]{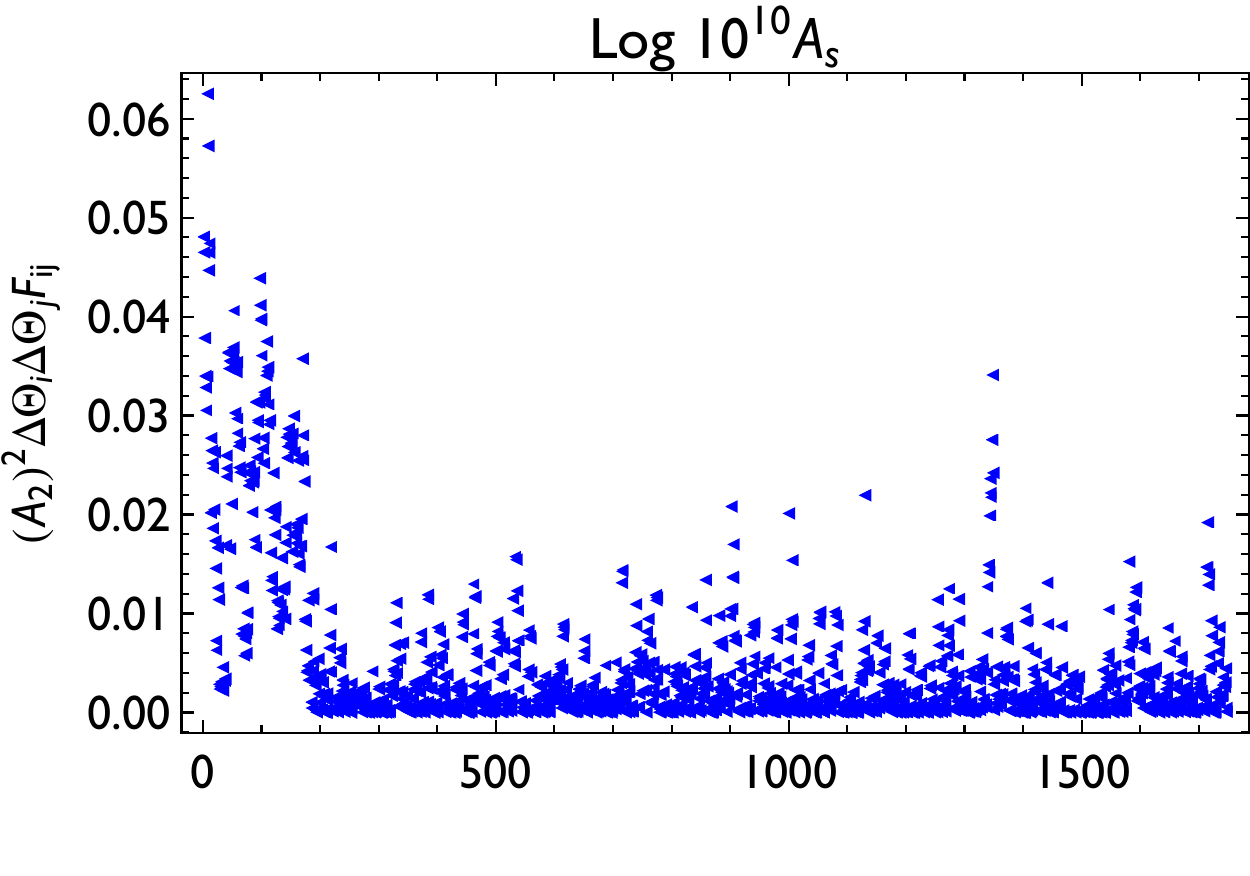}}\qquad
  \subfigure{\includegraphics[width=0.45\textwidth]{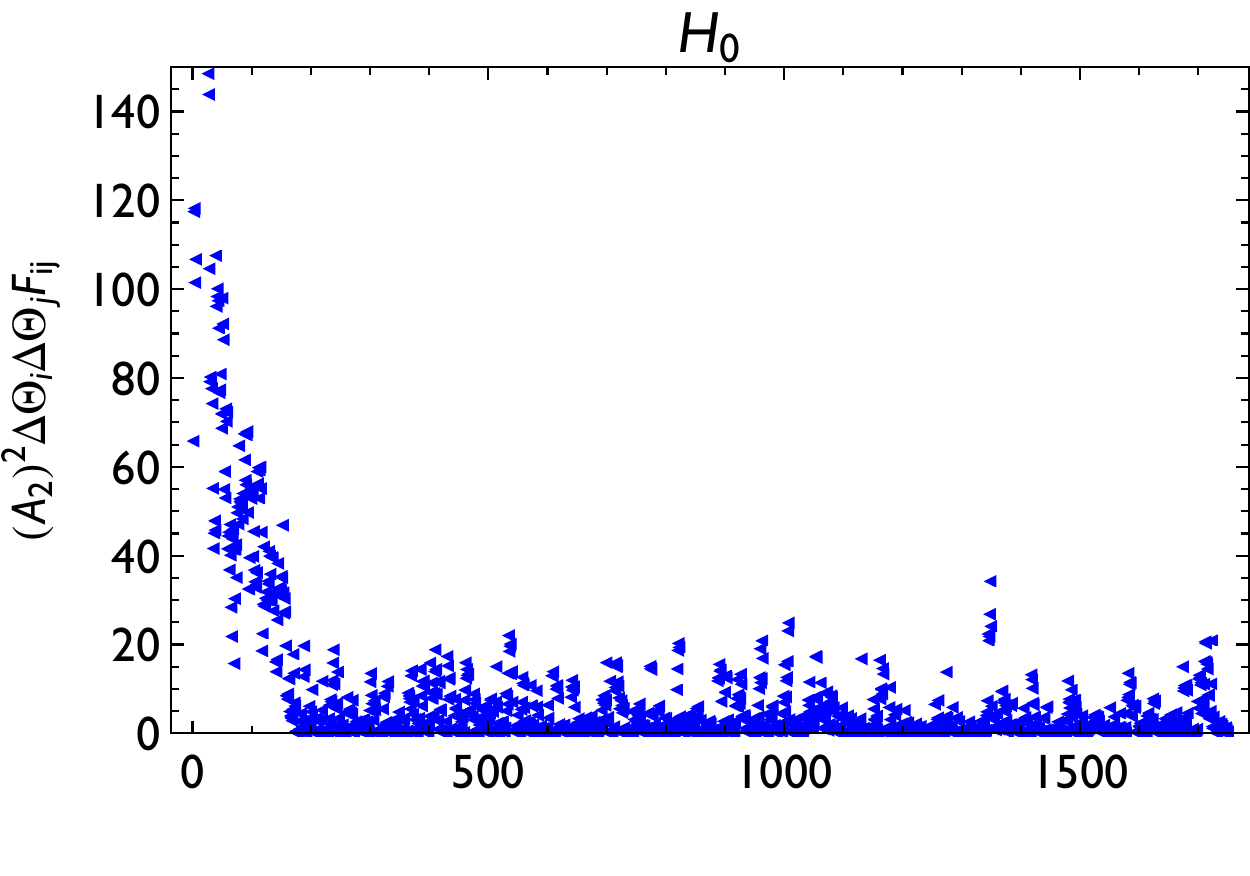}}
  \caption{The expression $A^2F_{ij}\Delta \Theta_i \Delta \Theta_j$ for (one of the 4) chains(s) for the best-fit $\omega_1=212.8$. It is clear from these plots that the cosmological parameters $\Omega_b h^2$, $\Omega_{dm}h^2$ and $H_0$ do not satisfy the bound if $\epsilon\leq\mathcal{O}(1)$ for most parameter values in the chains. It tells us that if the signal is real, we should expand to higher order and check if those cosmological parameters are either biased or have a smaller error. }
  \label{fig:fij}
\end{figure*}

We used the best-fit WMAP9  parameters to generate spectra in log space and in linear space. For log spaced oscillations we consider $10<\omega_1 < 250$, with a total of 1201 steps in frequency space (i.e. resolution of $\Delta \omega_1 = .2$). The improvement compared to no modulations is shown in Fig.~ \ref{fig:wmapchi}.  Clearly there are several frequencies that improve the fit, most remarkable around the frequencies identified earlier by \cite{2013arXiv1303.2616P} for log spaced oscillations. Unlike that work, our best-fit improvement is $\-2\Delta \log \mathcal{L}\sim15$. We investigated this difference in detail and we attribute the difference to them using primordial spectra computed directly from the inflaton potential, compared to our analysis using a approximated template. The best-fit parameters are given in Table \ref{tab:bestfitlogwmap9}. The best-fit has a large amplitude ($A=0.27$). We compute Eq.~\eqref{eq:fisherbound} for all $\Lambda$CDM parameters. They are shown in Fig.~\ref{fig:fij}. As expected, for such large amplitude and at these high frequencies, we expect that if this oscillations is real, we can in fact induce valuable information from the sampling of the transfer function (we can reduce the error bar on the cosmological parameters $\Omega_b h^2$, $\Omega_{dm}h^2$ and $H_0$ significantly).

\begin{table*}
\centering
\begin{tabular}{| l | c | c | c | c | c | c | c | c | r |}
\hline\hline 
Parameter & $\Omega_b h^2$ & $\Omega_c h^2$& $\tau$ & $H_0$ & $n_s $ &$\log 10^{10} A_s$ & $A_2/B_2$ & $\phi_1/\phi_2$ \tabularnewline
\hline 
Best-fit (log) & $0.022446$& 0.11506 & 0.08425 & 69.08 & 0.9688 &3.19 & 0.2705 & -0.48704  \tabularnewline
Best-fit (lin) & $0.022542$& 0.11264 & 0.08436 & 70.04 & 0.9718 &3.17 & 0.2707 & 2.01  \tabularnewline
\hline
\end{tabular}
\caption{best-fit parameter values for $\omega_1= 212.8$ with $-2 \Delta \log \mathcal{L}\simeq -15$ and  $\omega_2= 7500$ with $-2 \Delta \log \mathcal{L}\simeq -16$. Note that the best-fit amplitudes are almost equivalent.}
\label{tab:bestfitlogwmap9}
\centering
\end{table*}

\begin{figure}[htbp] 
   \centering
   \includegraphics[width=3.2in]{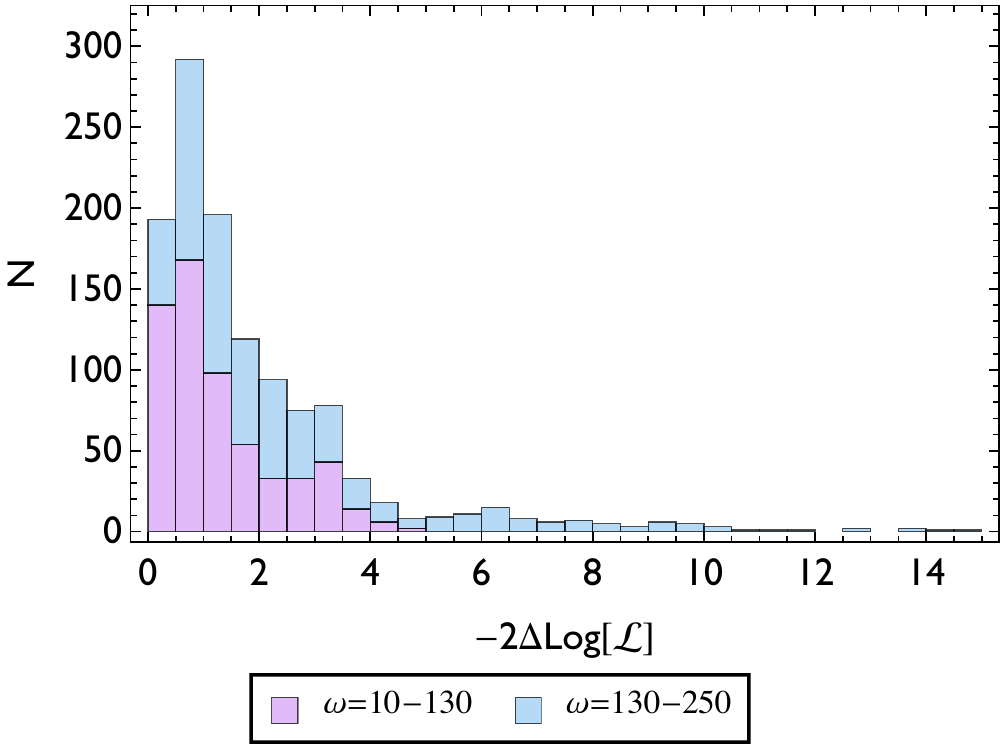} 
   \caption{ Histogram distribution of improvement in the likelihood. We made an (arbitrary) split in frequencies, to show that most improvements are at relatively high frequency. }
   \label{fig:wmaphisto}
\end{figure}

Fig.~\ref{fig:wmaphisto} shows the distribution of best-fit amplitudes as a function of $\Delta \chi^2$, where for comparison we split up the bins into (arbitrary) low frequency and high frequency, overall showing that for WMAP9  data, the low frequencies are constrained better than the high frequencies. In the analysis of fiducial Planck-like data earlier, we found that  {\it of the model is the correct model} simulations have shown that we expect an improvement of $-2\Delta \log \mathcal{L}  \sim -30 $ (for $\omega_1 \sim 210$) with an amplitude $A_2=0.1$, with Planck-like data. If in the WMAP9  data we are actually fitting the correct model (as in log-spaced oscillations), the improvement we find now $-2\Delta \log \mathcal{L}  \sim -15 $ is relatively small. We will  further comment on this in \S\ref{discussion}.

\begin{figure}[htbp] 
   \centering
   \includegraphics[width=3.4in]{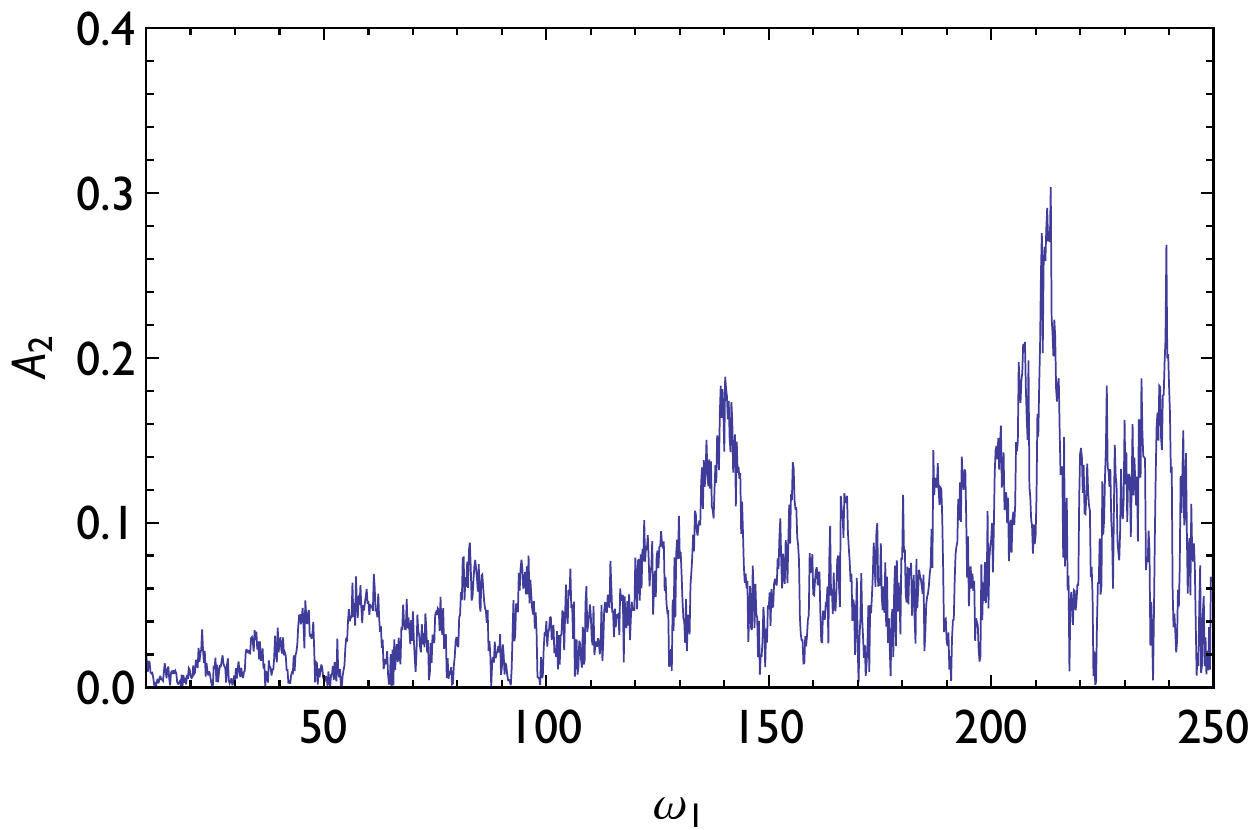} 
   \caption{The best-fit amplitude $A_2$ versus the frequency. Improvement of the fit is strongly correlated with the amplitude of the correction, as expected.}
   \label{fig:wmapamplitude}
\end{figure}

We also plot the distribution of best-fit amplitudes as a function of improvement of fit in Fig.~\ref{fig:wmapamplitude}.

\begin{figure}[htbp] 
   \centering
   \includegraphics[width=3.4in]{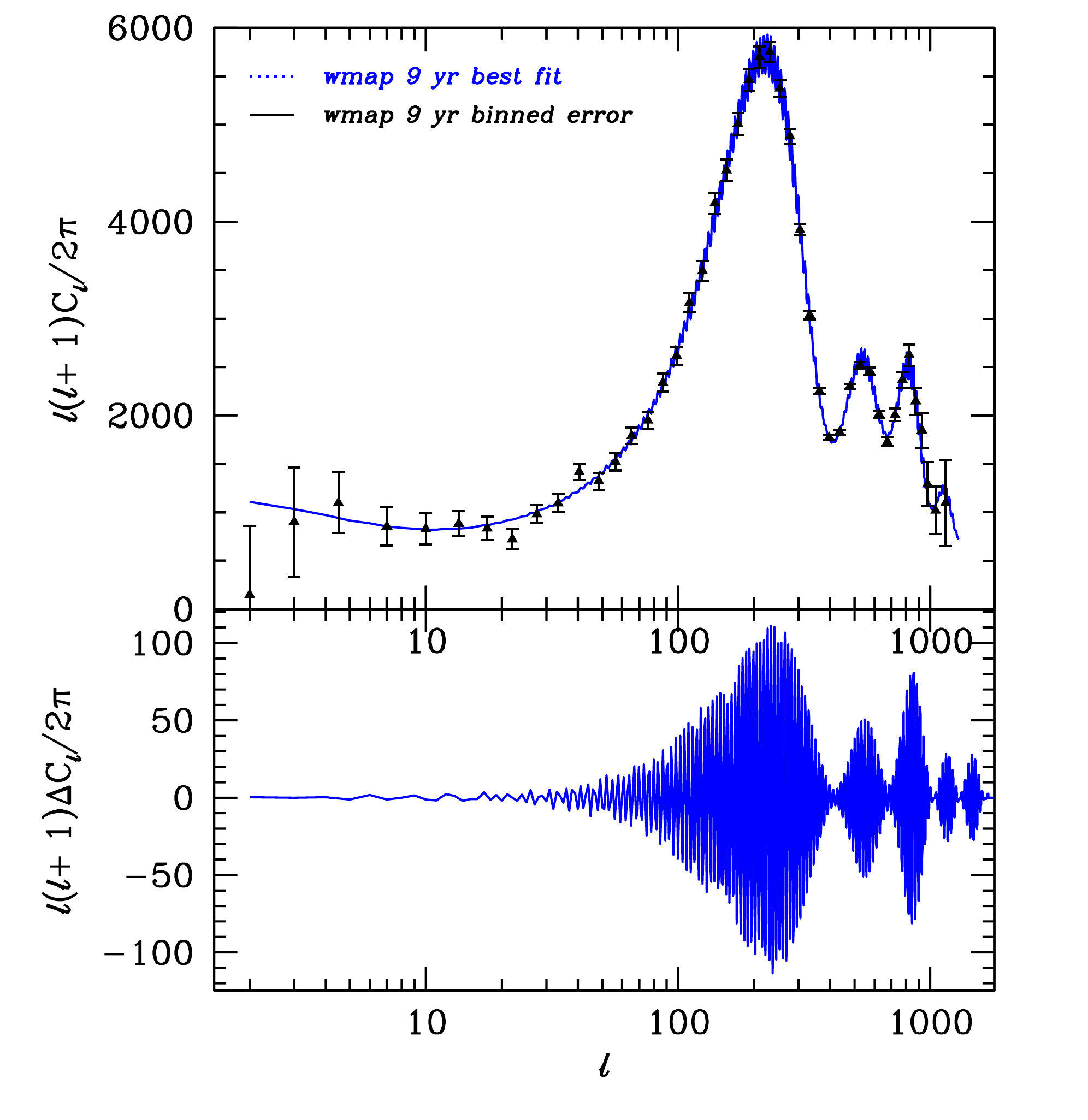} 
   \caption{The best-fit log spaced spectrum given WMAP9  yr data, plotted together with the residual and the $\Lambda$CDM covariance errors. }
   \label{fig:wmapbestfit}
\end{figure}

There are two possible explanations which could cause a large correction with a relatively small improvement of fit. The first possibility is that this is simply the statistical fluke (in our companion paper we will investigate this possibility by looking at similar oscillation in Planck). This is the most logical explanation, given that the improvement is small and we do not see a similar structure around the best-fit as we find in the fiducial data analysis. 

The second option is that there is an oscillation, but the template we are using is not sufficient to resolve the oscillations entirely, only recovering part of the signal through a mapping into log spaced oscillations. For example, one could image an inflationary model (e.g. with multiple axion) causing log spaced oscillations and features through bends in turns in field space. This could lead to resonance between the various primordial components and would make analysis very difficult (and evidence even harder to qualify), but it could explain a partial fit and therefore an improved likelihood with a relatively large amplitude?

\subsection{Linear-spaced oscillations}

\begin{figure}[htbp] 
   \centering
   \includegraphics[width=3.4in]{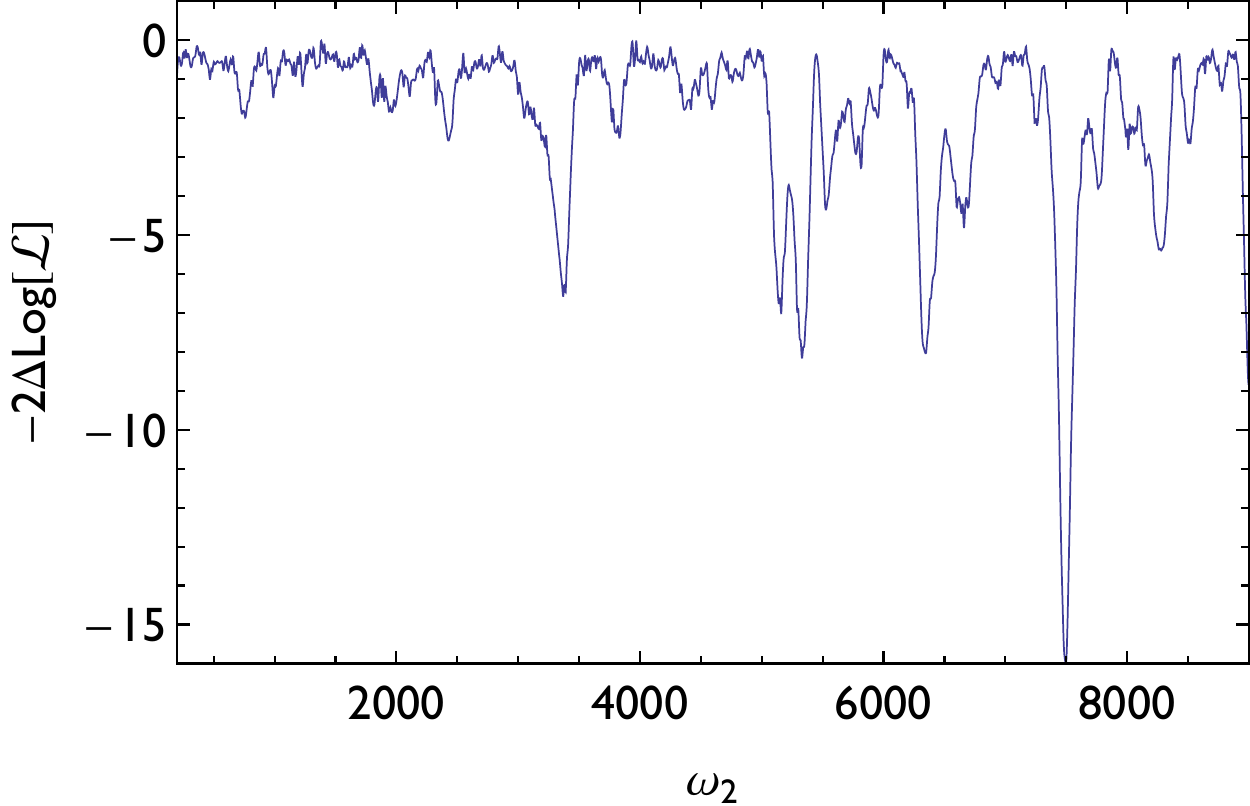} 
   \caption{The improvement of fit for 881 frequencies in WMAP9  data. }
   \label{fig:wmapchi2}
\end{figure}

\begin{figure}[htbp] 
   \centering
   \includegraphics[width=3.2in]{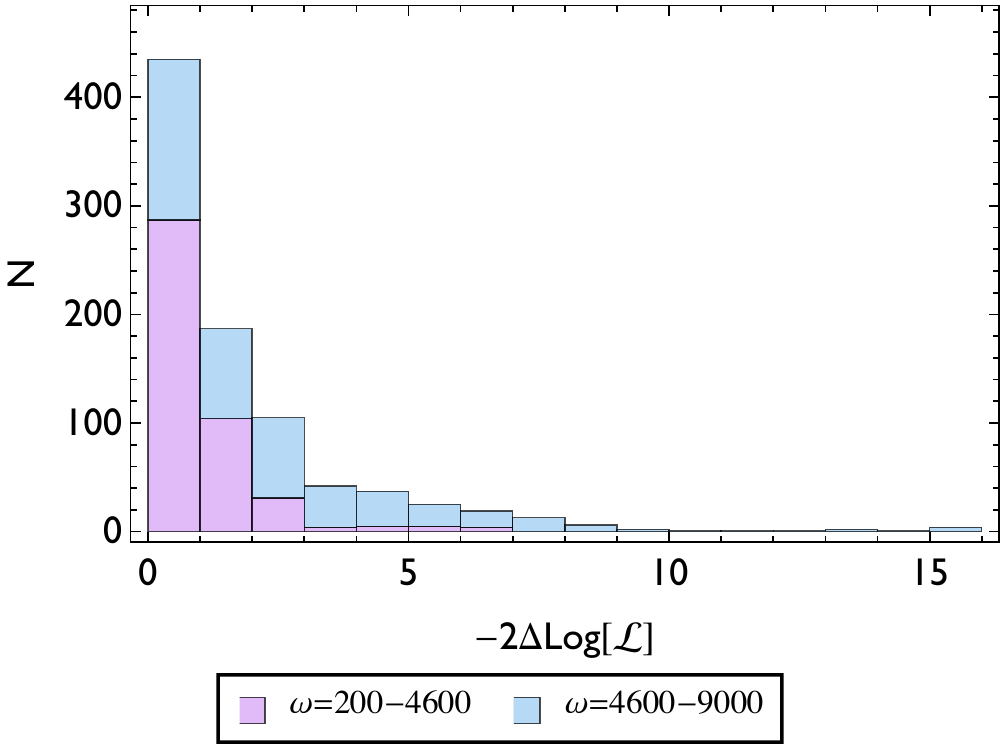} 
   \caption{Histogram distribution of improvement in the likelihood for linear spaced oscillations.  Again, the largest improvements are at relatively high frequency. }
   \label{fig:wmaphistol}
\end{figure}

For linear spaced oscillations we consider much higher frequencies between $200\leq \omega_2 \leq 9000$ given the suppression of the primordial frequency through projection. With a step width of $\Delta \omega_2 =10$, we analyze a total of 881 steps.  Fig.~\ref{fig:wmapchi2} shows several frequencies that lead to an improved fit over no oscillations. In particular we identify a peak $\omega_2 = 7500$, with $-2\Delta \log \mathcal{L}  \sim -16 $, similar to the best improvement for log spaced oscillations. The best-fit has an amplitude of $B_2=0.27$ and a phase $\phi_2 = 2.01$ (see Tab. \ref{tab:bestfitlogwmap9}). Fig. \ref{fig:wmaphistol} shows a histogram of the improvements found for the 881 sampled frequencies.

\begin{figure}[htbp] 
   \centering
   \includegraphics[width=3.4in]{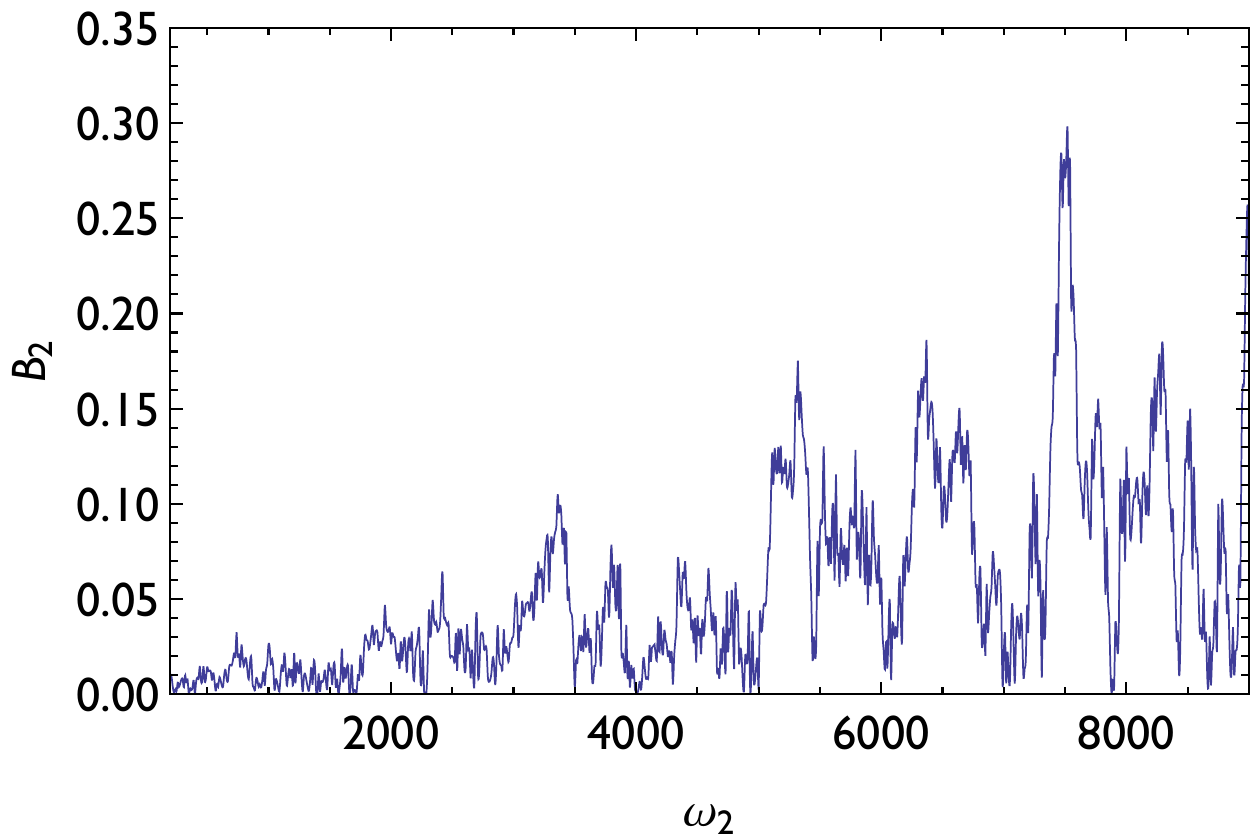} 
   \caption{The best-fit amplitude $B_2$ versus the frequency.}
   \label{fig:wmapamplitude2}
\end{figure}
\begin{figure}[htbp] 
   \centering
   \includegraphics[width=3.4in]{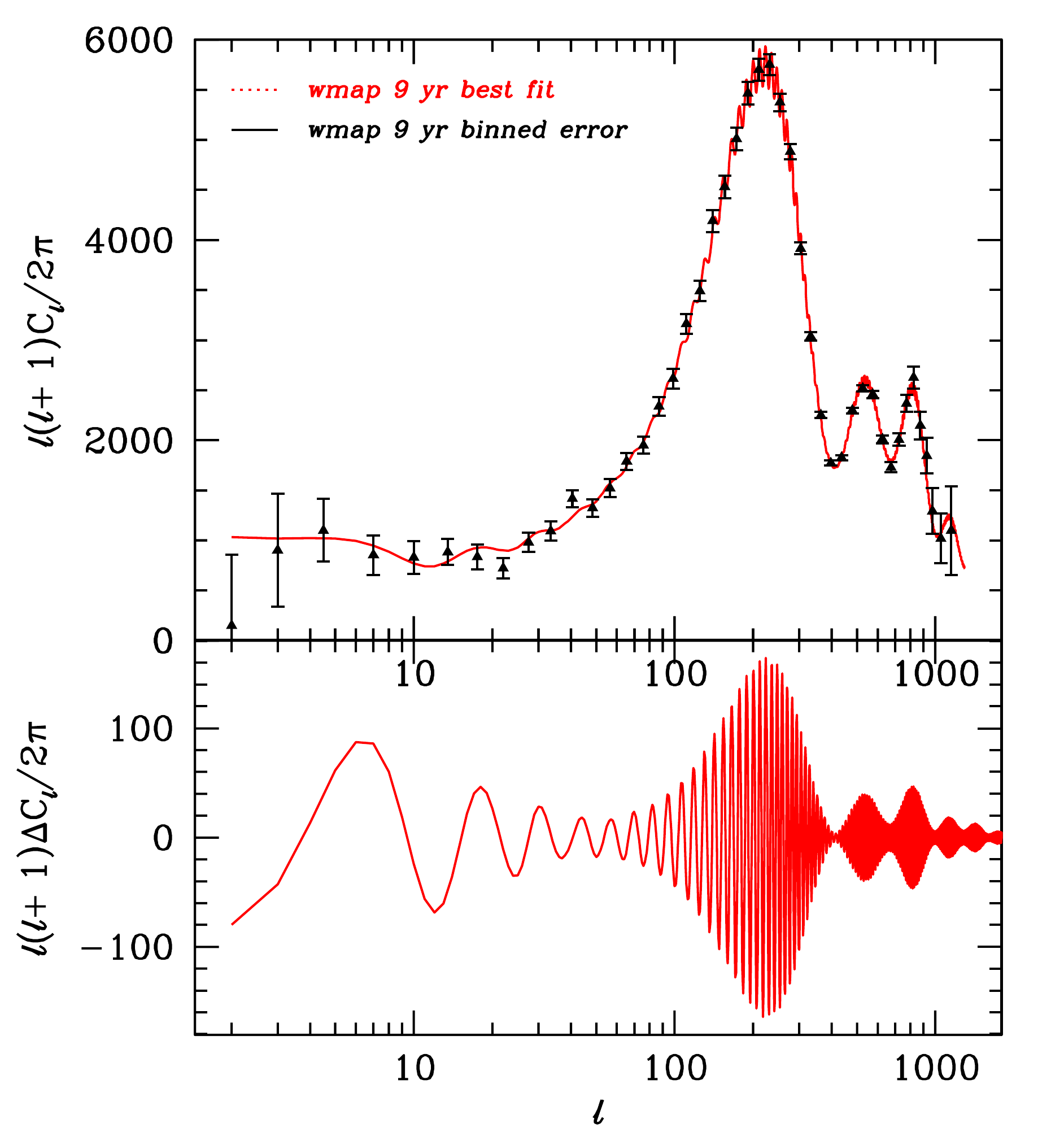} 
   \caption{The best-fit spectrum for linear spaced oscillations given WMAP9  yr data, plotted together with the residual and the $\Lambda$CDM covariance errors. }
   \label{fig:wmapbestfit}
\end{figure}

\section{Discussion} \label{discussion}

\subsection{Model selection statistics}

Does a spectrum with oscillations provide a better fit to the data?
For each model, we have found oscillation frequencies that provide  a better fit of the data than the no oscillation model.
However, the improvement in the fit is smaller than the improvement found in simulations
for input models with oscillations. Since the purpose of this paper is to show the methodology works on Planck-like data, we have focussed our tests on Planck-like simulations. In this section we will apply several information criteria that weight each model according to the number of data points fitted as well as the number of free parameters. An obstacle in actually weighting the likelihood of each model is set by the fact that although we fit each frequency independently (we run chains for a fixed frequency), in principle there are only 2 primordial spectra : one with and one without oscillations. In other words, should we compare between these two models or should we compare between frequencies, sampled in each oscillator model? For this purpose we can consider each frequency a different model (which would set the number of unknown parameters from 6 to 8 and the number of models to $n$, with $n$ the number of trials).

The Bayesian evidence methodology provides a framework for rigorously answering this significance of the oscillations.  
However, evaluating the Bayesian evidence requires specifying the priors.  These
depend on the underlying physical model and differ for each of the physical mechanism for generating oscillations in the spectrum. Here, we simply present the information criteria as a general weight to the likelihood of the data given the model, and for those interested we will make the data publicly available such that each model of interest can be tested individually. We refer the reader to \citep{2013arXiv1303.2616P} for an evidence-based analysis of a specific oscillation model. 

We will consider three different information criteria (for a recent discussion see e.g. Refs. \citep{2007MNRAS.377L..74L} and \citep{2013MNRAS.tmp.1367M}) The first one of these is the Akaike Information Criteria (AIC) is given by
\be
AIC = -2 \ln \mathcal{L}_{\mathrm{max}} + 2k
\ee
with $k$ the number of parameters. There is a punishment for adding more parameters, through the term $2k$. Other have increased the punishment (i.e. over fitting) by changing the $2k\rightarrow 3k$, which is referred to as the Kullback information criterium. The evidence is generally considered weak if the difference $AIC_1-AIC_2$ is less than 2, and strong if this difference $>5$. In the case we consider each model independently for each frequency we searched for, we can also define the Akaike weight 
\be
\mathcal{L}(M_{a}) = \frac{\exp(-AIC_a/2)}{\prod_N \exp(-AIC_1/2)\ldots \exp(-AIC_N/2)} 
\label{eq:AICweight}
\ee
which naturally takes into account the look-elsewhere effect. 

The Bayesian Information Criterium takes into account the number of degrees of freedom (or fitting points) and the penalty of over-fitting is proportional to the log of that, i..e 
\be
BIC = -2 \ln \mathcal{L}_{\mathrm{max}}+  k \log n
\ee
with $n$ the number of data points. Since $n$ is equivalent for all our models (including $\Lambda$CDM) we will be only concerned with the difference in  $-2 \ln \mathcal{L}_{\mathrm{max}}$ and the number of parameters for the $AIC$ and $BIC$ criteria, while for the Bayesian information criterium we also need to take into account the number of data points, which for WMAP is equivalent to $l_{max} = 1200$.

\begin{table*}
\centering
\begin{tabular}{| l | c | c | r |}
\hline\hline 
Information& WMAP9  log  &  WMAP9  lin & \tabularnewline
\hline 
AIC & 11  & 12 &  \tabularnewline
KIC & 9 & 10 &  \tabularnewline
BIC & 1& 2 &  \tabularnewline
\hline
\end{tabular}
\caption{Several information criteria. Here we assume $k=2$, and $n=1200$. }
\label{tab:INFOCR}
\centering
\end{table*}

\begin{figure}[htbp] 
   \centering
   \includegraphics[width=3in] {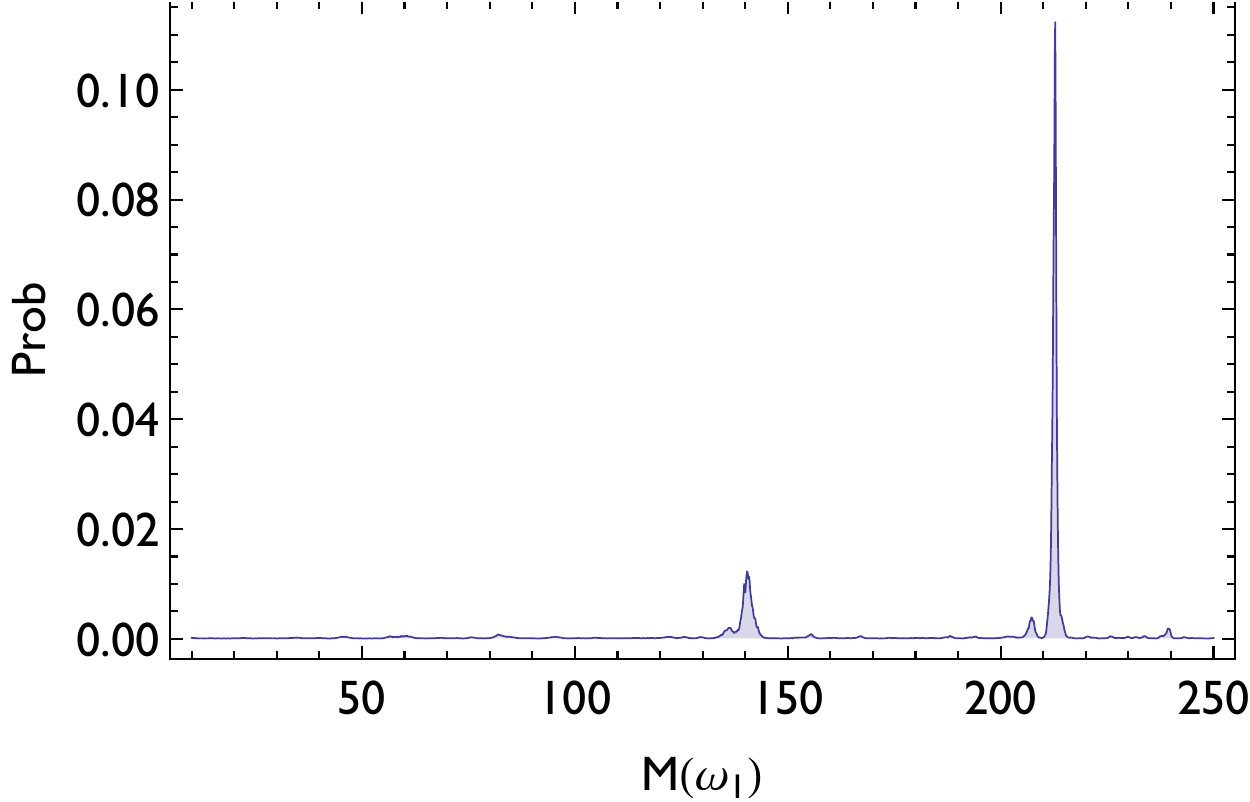} 
   \caption{The Akaike weight as defined in \protect{\eqref{eq:AICweight}}. We treat every discrete frequency investigated as a distinct model. This weights the fact that we consider so many different frequencies and suppresses the probability of any given find. Note that this distribution is a measure of improvement (in the set of 1201 trials), therefore if we set a detection limit at $3\sigma$ (roughly assuming the distribution of improvements is a Gaussian as shown in Fig.~\ref{fig:wmaphisto}, $P = 0.997$), none of the best-fit oscillations can be considered a detection. }
   \label{fig:Aweight}
\end{figure}

\begin{figure}[htbp] 
   \centering
   \includegraphics[width=3in] {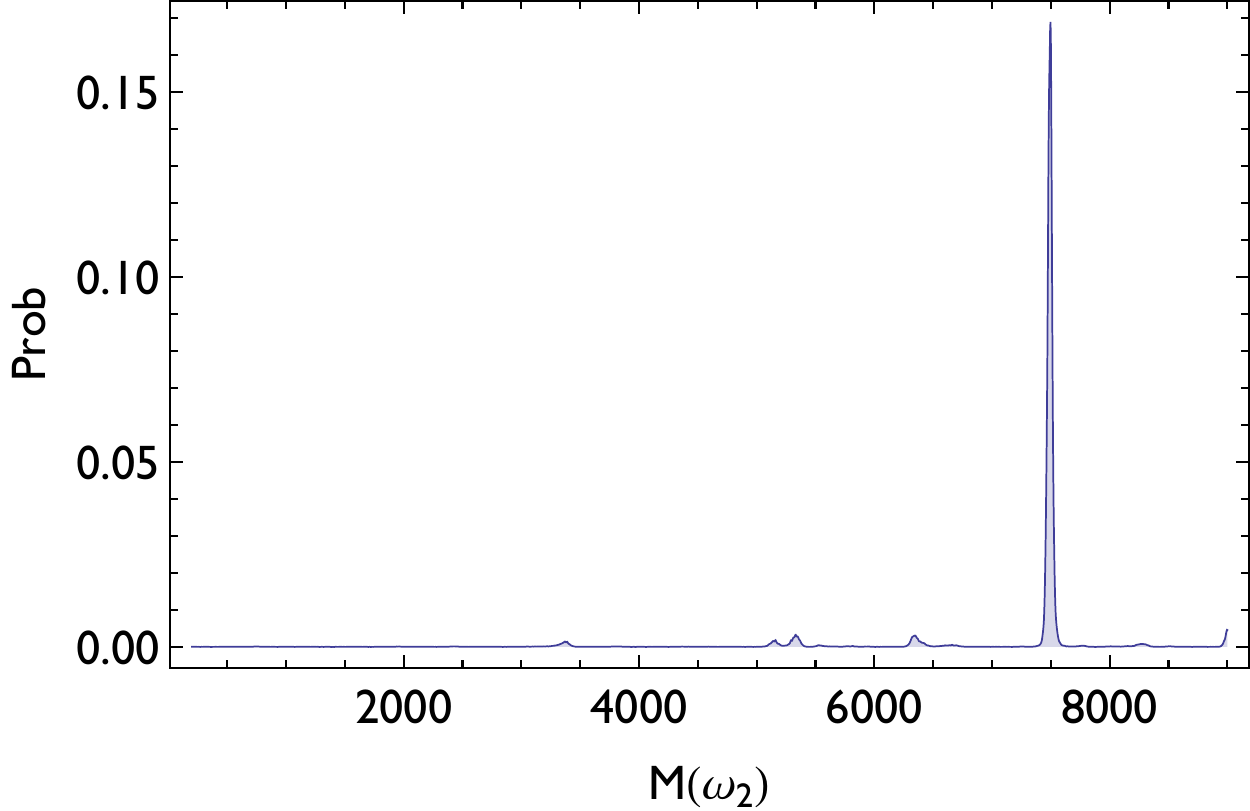} 
   \caption{The Akaike weight now for linear oscillations. The peak probability is higher because of the fewer trials (881 versus 1201) and fewer peaks. }
   \label{fig:Aweight2}
\end{figure}

We show the results in Table~\ref{tab:INFOCR}. In Fig.~\ref{fig:Aweight} and Fig.~\ref{fig:Aweight2}  we show the Akaike weights of both the log and linear model. 
It is clear that each information criteria could lead you to either believe there is sufficient evidence (Akaike and Kullback) in favor of the best-fit amplitude, as well as no evidence (Bayesian). 

\subsection{Monte Carlo}
\begin{figure}[htbp] 
   \centering
   \includegraphics[width=3.2in]{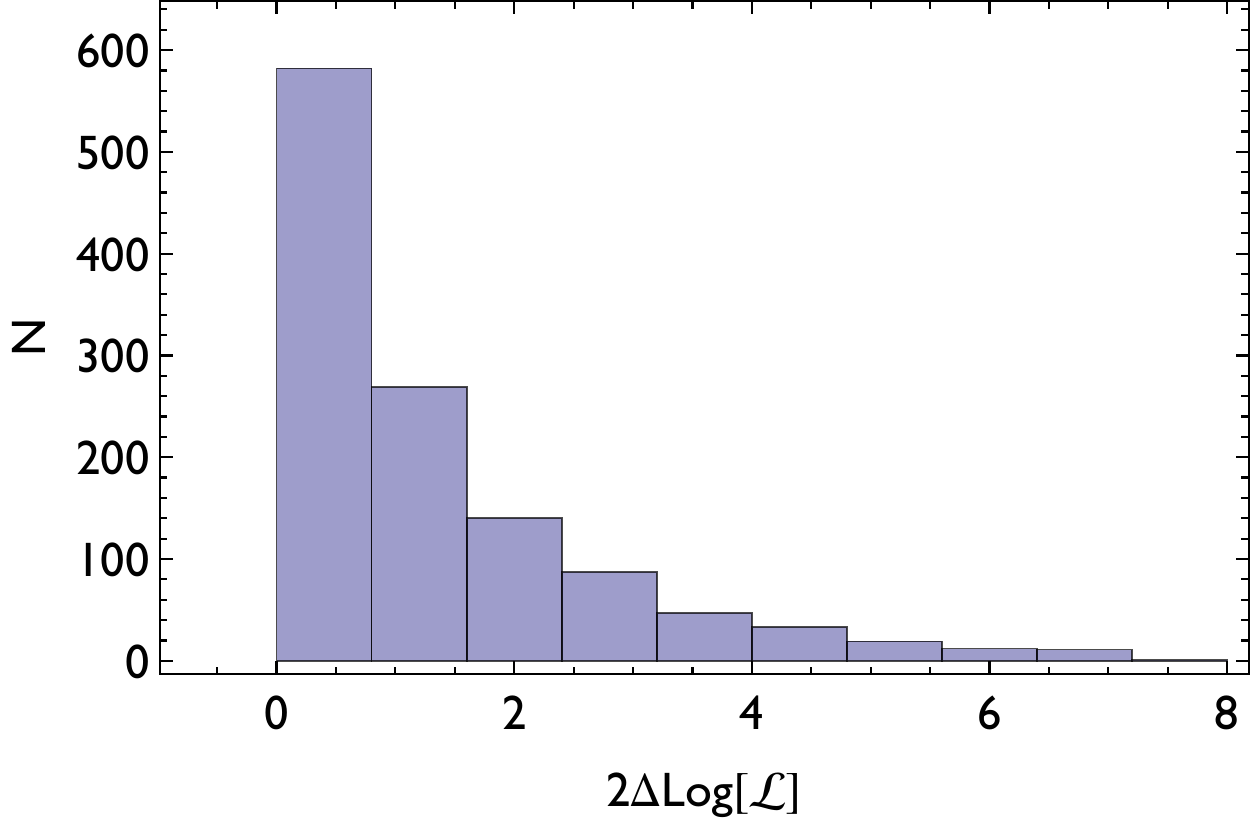} 
   \caption{Histogram distribution of improvements for log spaced oscillations in the likelihood for simulated nul data, with a fixed random seed for the noise.}
   \label{fig:simloghisto}
\end{figure}

To  further investigate possible significance of the 2 peaks in WMAP and possible features in Planck (see companion paper), we ran two additional tests. First, we generated random Planck-like data as before with no signal. We set $\ell_{\mathrm{max}}=2000$ and use the same noise statistics as before. For log spaced oscillations we ran 1201 frequencies and a histogram of the improvement of fit is shown in Fig.~\ref{fig:simloghisto}. We find that with a fixed random seed, the maximum improvement of the likelihood is $2 \Delta\log \mathcal{L} \sim 8$. The best-fit amplitude $A_2\sim 0.045$, which suggests that indeed fluctuations in the noise can at least mimic log space oscillations up to a fluctuation of $A_2\lesssim .05$, which explains the observation that for fluctuations below this level any true primordial signal will most likely become entangled with fluctuations in the noise. Recall that projection suppresses the observed amplitude of the fluctuations, and the largest amplitude at low frequencies ($\omega_1<100$) for this noise seed shows $A_2\leq .02$ with $2 \Delta \log \mathcal{L}\leq 6$. We find $\bar{A}_2 = 0.013$ with a standard deviation of $0.008$, which indeed suggests amplitudes $A_2\sim 0.1$ and below are most likely noise or at least are noise dominated. 

\begin{figure}[htbp] 
   \centering
   \includegraphics[width=3.2in]{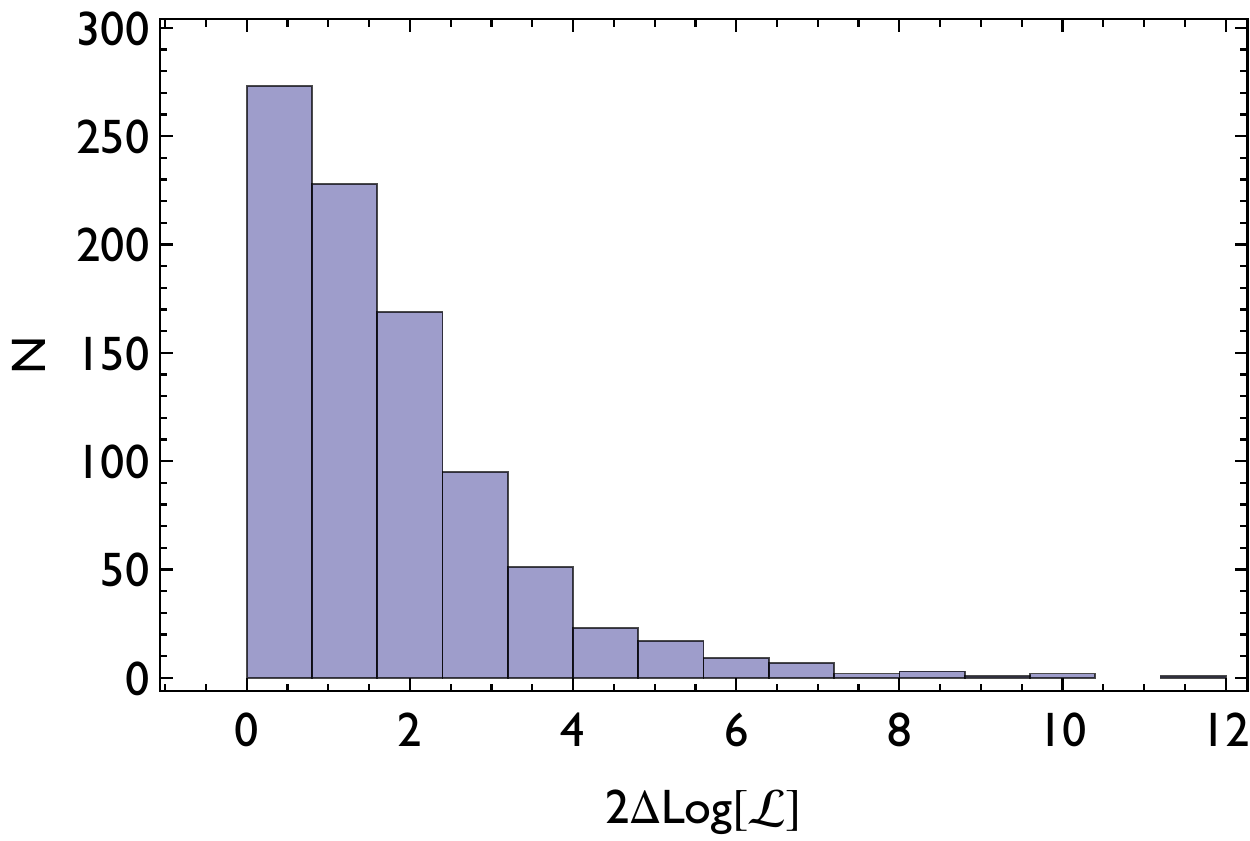} 
   \caption{Histogram distribution of improvements for linear spaced oscillations in the likelihood for simulated nul data, with a fixed random seed for the noise. For the same noise seed, we find improvements that are better. }
   \label{fig:simlinhisto}
\end{figure}

Likewise, we performed an analysis using the linear spaced oscillations over the same frequency range as we used to analyze the WMAP data. The histogram of the improvement is shown in Fig.~\ref{fig:simlinhisto}. We used the same/different null signal maps as we used for the log space analysis. We find that the best-fit improvement is $2\Delta \log \mathcal{L}\sim 12$ with the biggest improvements at high frequencies. The mean fitted amplitude is $\bar{B}_2 = 0.024$ with a standard deviation of $0.015$. 

\begin{figure}[htbp] 
   \centering
   \includegraphics[width=3in] {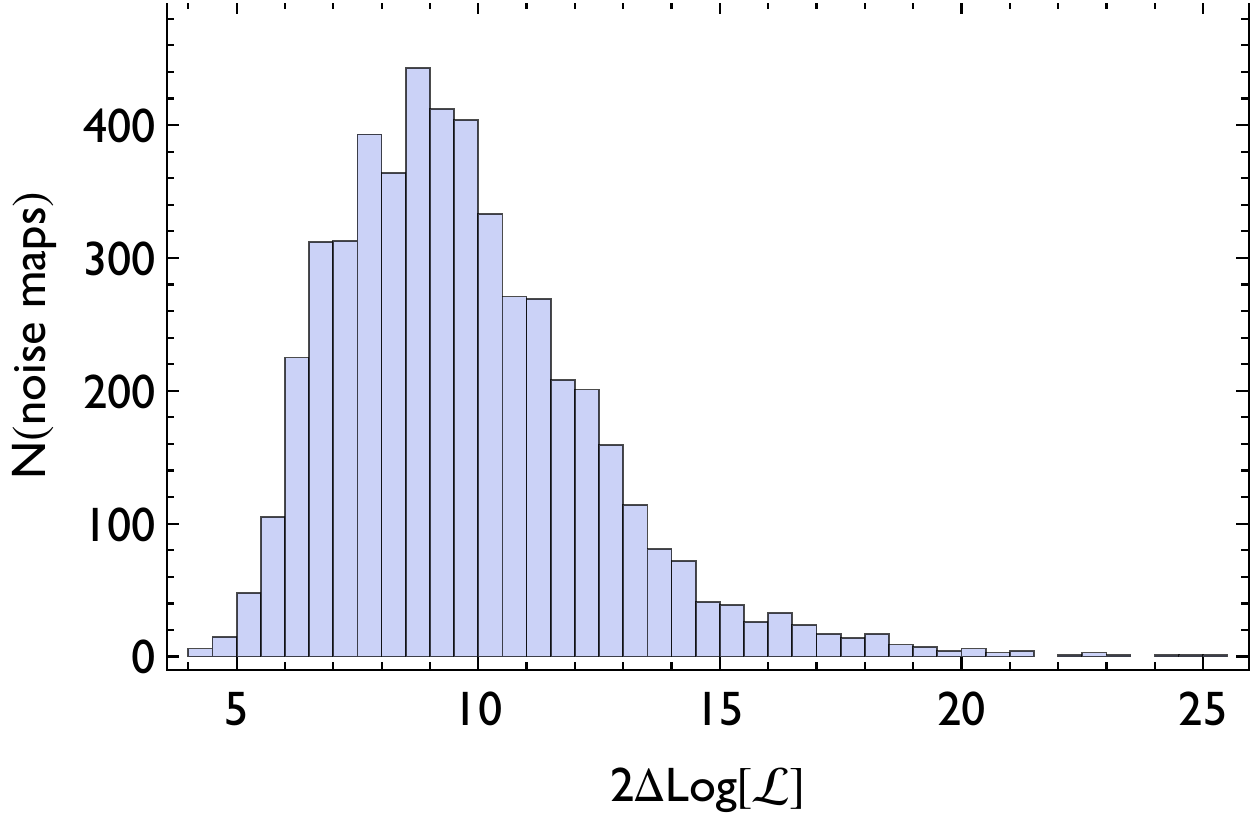} 
   \caption{The distribution of $2\Delta \log \mathcal{L}_{\mathrm{max}}$ for log spaced oscillations. We used Gaussian noise and ran a grid with the following spacing $-\pi \leq \phi_1 \leq \phi$ ($\Delta \phi_1 = \pi/2$),  $10 \leq \omega_1 \leq 250$ ($\Delta \omega_1 = 1$) and $0 \leq A_2^{\mathrm{eff}} \leq 0.06$ ($\Delta A_2^{\mathrm{eff}}= 0.005$), where the effective amplitude is the amplitude set after projection. }
   \label{fig:mclog}
\end{figure}

Secondly\footnote{The idea for this test was suggested by Raphael Flauger (private communication). His results will be published in a forthcoming paper. Something very similar was done for a free-form power spectrum in \cite{2010JCAP...04..010H}}, given the improvements we found in a single run, we are interested in what the typical maximum improvement is due to a possible fitting of the noise (i.e. for the mock data above $8$ and $12$ respectively). To investigate this we ran a large set of simulations (5000), performing a similar analysis. In order to speed up calculations we simplified our search significantly. We generated mock data with a single channel and set $\ell_{\mathrm{max}} =500$. We coded a simple $\chi^2$ fitting, were we first fix the primordial amplitude $A_s$ to the best-fit. After that we run a grid, varying the amplitude, the phase and the frequency, with sufficient step size. Each analysis is performed on a data set with random noise, drawing from a normal distribution (Gaussian noise), including cosmic variance. We store the best-fit of each run in a data file. The results are shown in Fig. \ref{fig:mclog} (log-spaced), and Fig.~\ref{fig:mclog} (linear spaced) which shows a distribution of improvements. This simple analysis shows that one typically expects $2\Delta \log \mathcal{L}\sim \mathcal{O}(10)$ (the mean for the log/linear is 9.8/10.2 with a maximum of 25.5 and 24.9 respectively). We find that the improvement for log spaced oscillations is in the 96 percentile and linear spaced oscillation in the 74 percentile.  {\it This suggests that the improvements we find can be completely explained by a fitting of the noise}. 

\begin{figure}[htbp] 
   \centering
   \includegraphics[width=3in] {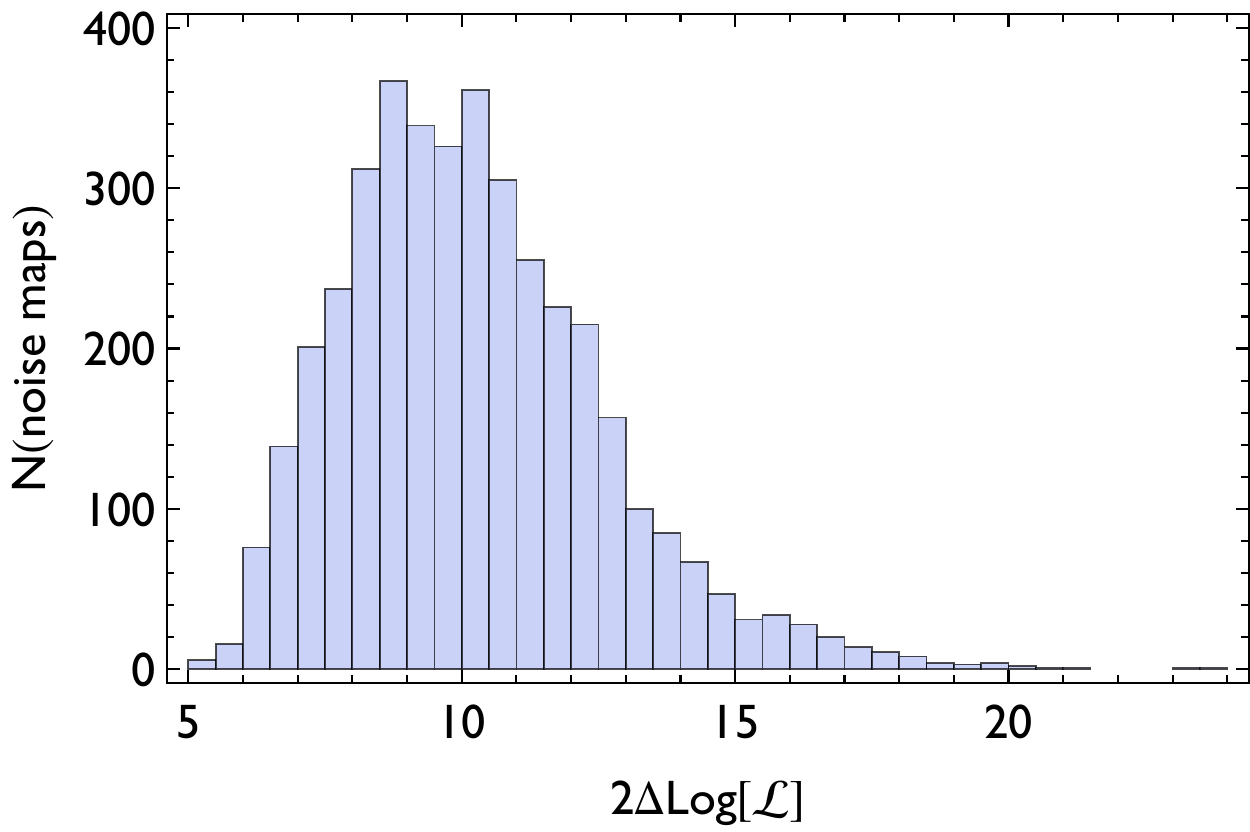} 
   \caption{The distribution of $2\Delta \log \mathcal{L}_{\mathrm{max}}$ for linear spaced oscillations. We used the following spacing $-\pi \leq \phi_1 \leq \phi$ ($\Delta \phi_1 = \pi/2$),  $200 \leq \omega_2 \leq 9000$ ($\Delta \omega_2 = 40$) and $0 \leq B_2^{\mathrm{eff}} \leq 0.06$ ($\Delta B_2^{\mathrm{eff}}= 0.005$). }
   \label{fig:mclin}
\end{figure}
\section{Conclusion} \label{conclusion}

In this paper, we presented a simple method that enables the rapid computation of the angular power spectrum even when the  primordial power spectrum has multiple oscillatory features. The method assumes the amplitude of the oscillatory part of the primordial power spectrum is small, thus, we can expand that the spectrum in a Taylor series. We expand up to any order we want, with little compromise on speed. We have shown that for Planck-like data, we only need to expand to lowest order in the transfer functions to get accurate results, as long as the anisotropy power due to the oscillations is only a fraction of the total power. We applied our code to simulated data, and found that we were capable to recover fiducial oscillations as long as the amplitude is greater than a few $\%$ of the primordial amplitude, although projection increases the amplitude at which a potential signal can be recovered at higher frequencies. 

In this paper we tested our code on WMAP9  year data release. For log-spaced oscillations we recovered 2 frequencies earlier identified in \citep{2013arXiv1303.2616P}. For linear spaced oscillations we were able to identity one frequency that gives a comparable improvement of fit. Both best-fitted frequencies (log and linear spectra) are large with many oscillations in the multipole domain ($\ell_{\mathrm{max}}=1200$) and because of projection the primordial amplitude is rather large with (interestingly) $A_2 = B_2 \simeq0.27$ as best-fit values. 

In order to address the potential significance of these findings we derived several familiar information criteria used in the literature, which shows that the significance of
these features.  We do not find  compelling evidence for features in the WMAP9 data. Further investigation by means of a Monte Carlo of fiducial data without oscillations shows that noise can easily produce a similar improvements of fit. Foremost, we ran a full pipeline analysis of our code, with a single seeded null map, showing that an improvement of the fit due to a fit to the noise leads to $2 \Delta\log \mathcal{L} \sim 10$ . We also run a simplified analysis with Planck-like data, generating a total of 5000 spectra for each model. Applying a $\chi^2$ fitting showed that $2 \Delta\log \mathcal{L}$ of $\mathcal{O}(10)$ are expected. In fact, $2 \Delta\log \mathcal{L} \geq 20$ are not uncommon. Although this analysis is extremely simplified, with only one channel and $\ell_{\mathrm{max}}$ set to 500, it suggest that any fit that does not produce an improvement $>20$ in $\chi^2_{\mathrm{eff}}$, carries a large risk of being the result of fitting oscillatory features to either noise or cosmic variance in the spectra.

This conclusion is supported by simulated maps that contain an oscillatory signal. Here we found that simulations with a signal typically produce a (much) larger improvement of the oscillatory correction is more than a few percent of the primordial amplitude. This could suggest two things: either the model we are considering is simply not the correct model or we are fitting the noise. In the first of these two possibilities, the primordial signal can be due to resonance type effects, but the model applied is wrong. We are getting a better fit, but additional effects need to taken into account in order to get a true improvement of fit. For example, there could be multiple axions or perhaps the feature is localized. Although an envelope shape of the feature can be implemented, multiple oscillations are much harder to test.  It was already shown by \cite{2013arXiv1303.2616P} that the log spaced oscillations do not lead to a gradual improvement of fit as a function of $\ell$. If the oscillation is a truly present, this is generally what we expect. For linear space oscillations, theoretical models typically predict a localized nature, so a local improvement can not be considered as counter evidence. We will investigate the $\ell_{\mathrm{max}}$ dependence in our companion paper.  Moreover if the features seen in the WMAP9 data were due to oscillations in the primordial
spectrum, then their significance should increase with the additional of more data (Planck).

While we were carrying out these investigations, other groups have made very similar attempts to look for resonant features in the CMB data \citep{2013Flaugera}. Since those codes work differently, we believe that our results are complementary. They apply the use of the multi nest sampler which allowed them to do an evidence check. Ideally, combining the two could lead to an extremely efficient code (going to high frequency in a single MCMC run). We look forward to implementing such improvements in our current pipeline. \\

\section*{Acknowledgments}

The authors would like to thank Guido D'Amico, Fabian Schmidt, Renee Hlozek and Kendrick Smith for useful discussions and comments. P.D.M. would especially like to thank Raphael Flauger for very useful discussions. P.D.M is supported by the Netherlands Organization for Scientific Research (NWO), through a Rubicon fellowship. P.D.M. and D.N.S. are in part funded by the John Templeton Foundation grant number 37426. B.D.W.
acknowledges funding through the ANR Chaire d'\'Excellence,
the UPMC Chaire Internationale in Theoretical Cosmology, and NSF grant AST-0708849. \\
 
 \bibliography{paper1_final}

\begin{thebibliography}{35}%
\makeatletter
\providecommand \@ifxundefined [1]{%
 \@ifx{#1\undefined}
}%
\providecommand \@ifnum [1]{%
 \ifnum #1\expandafter \@firstoftwo
 \else \expandafter \@secondoftwo
 \fi
}%
\providecommand \@ifx [1]{%
 \ifx #1\expandafter \@firstoftwo
 \else \expandafter \@secondoftwo
 \fi
}%
\providecommand \natexlab [1]{#1}%
\providecommand \enquote  [1]{``#1''}%
\providecommand \bibnamefont  [1]{#1}%
\providecommand \bibfnamefont [1]{#1}%
\providecommand \citenamefont [1]{#1}%
\providecommand \href@noop [0]{\@secondoftwo}%
\providecommand \href [0]{\begingroup \@sanitize@url \@href}%
\providecommand \@href[1]{\@@startlink{#1}\@@href}%
\providecommand \@@href[1]{\endgroup#1\@@endlink}%
\providecommand \@sanitize@url [0]{\catcode `\\12\catcode `\$12\catcode
  `\&12\catcode `\#12\catcode `\^12\catcode `\_12\catcode `\%12\relax}%
\providecommand \@@startlink[1]{}%
\providecommand \@@endlink[0]{}%
\providecommand \url  [0]{\begingroup\@sanitize@url \@url }%
\providecommand \@url [1]{\endgroup\@href {#1}{\urlprefix }}%
\providecommand \urlprefix  [0]{URL }%
\providecommand \Eprint [0]{\href }%
\providecommand \doibase [0]{http://dx.doi.org/}%
\providecommand \selectlanguage [0]{\@gobble}%
\providecommand \bibinfo  [0]{\@secondoftwo}%
\providecommand \bibfield  [0]{\@secondoftwo}%
\providecommand \translation [1]{[#1]}%
\providecommand \BibitemOpen [0]{}%
\providecommand \bibitemStop [0]{}%
\providecommand \bibitemNoStop [0]{.\EOS\space}%
\providecommand \EOS [0]{\spacefactor3000\relax}%
\providecommand \BibitemShut  [1]{\csname bibitem#1\endcsname}%
\let\auto@bib@innerbib\@empty
\bibitem [{\citenamefont {{Starobinsky}}(1980)}]{1980PhLB...91...99S}%
  \BibitemOpen
  \bibfield  {author} {\bibinfo {author} {\bibfnamefont {A.~A.}\ \bibnamefont
  {{Starobinsky}}},\ }\href {\doibase 10.1016/0370-2693(80)90670-X} {\bibfield
  {journal} {\bibinfo  {journal} {Physics Letters B}\ }\textbf {\bibinfo
  {volume} {91}},\ \bibinfo {pages} {99} (\bibinfo {year} {1980})}\BibitemShut
  {NoStop}%
\bibitem [{\citenamefont {{Mukhanov}}\ and\ \citenamefont
  {{Chibisov}}(1981)}]{1981ZhPmR..33..549M}%
  \BibitemOpen
  \bibfield  {author} {\bibinfo {author} {\bibfnamefont {V.~F.}\ \bibnamefont
  {{Mukhanov}}}\ and\ \bibinfo {author} {\bibfnamefont {G.~V.}\ \bibnamefont
  {{Chibisov}}},\ }\href@noop {} {\bibfield  {journal} {\bibinfo  {journal}
  {ZhETF Pisma Redaktsiiu}\ }\textbf {\bibinfo {volume} {33}},\ \bibinfo
  {pages} {549} (\bibinfo {year} {1981})}\BibitemShut {NoStop}%
\bibitem [{\citenamefont {Guth}(1981)}]{PhysRevD.23.347}%
  \BibitemOpen
  \bibfield  {author} {\bibinfo {author} {\bibfnamefont {A.~H.}\ \bibnamefont
  {Guth}},\ }\href {\doibase 10.1103/PhysRevD.23.347} {\bibfield  {journal}
  {\bibinfo  {journal} {Phys. Rev. D}\ }\textbf {\bibinfo {volume} {23}},\
  \bibinfo {pages} {347} (\bibinfo {year} {1981})}\BibitemShut {NoStop}%
\bibitem [{\citenamefont {{Linde}}(1982)}]{1982PhLB..108..389L}%
  \BibitemOpen
  \bibfield  {author} {\bibinfo {author} {\bibfnamefont {A.~D.}\ \bibnamefont
  {{Linde}}},\ }\href {\doibase 10.1016/0370-2693(82)91219-9} {\bibfield
  {journal} {\bibinfo  {journal} {Physics Letters B}\ }\textbf {\bibinfo
  {volume} {108}},\ \bibinfo {pages} {389} (\bibinfo {year}
  {1982})}\BibitemShut {NoStop}%
\bibitem [{\citenamefont {{Planck Collaboration}}\ \emph
  {et~al.}(2013{\natexlab{a}})\citenamefont {{Planck Collaboration}},
  \citenamefont {{Ade}}, \citenamefont {{Aghanim}}, \citenamefont
  {{Armitage-Caplan}}, \citenamefont {{Arnaud}}, \citenamefont {{Ashdown}},
  \citenamefont {{Atrio-Barandela}}, \citenamefont {{Aumont}}, \citenamefont
  {{Baccigalupi}}, \citenamefont {{Banday}},\ and\ \citenamefont
  {et~al.}}]{2013arXiv1303.5082P}%
  \BibitemOpen
  \bibfield  {author} {\bibinfo {author} {\bibnamefont {{Planck
  Collaboration}}}, \bibinfo {author} {\bibfnamefont {P.~A.~R.}\ \bibnamefont
  {{Ade}}}, \bibinfo {author} {\bibfnamefont {N.}~\bibnamefont {{Aghanim}}},
  \bibinfo {author} {\bibfnamefont {C.}~\bibnamefont {{Armitage-Caplan}}},
  \bibinfo {author} {\bibfnamefont {M.}~\bibnamefont {{Arnaud}}}, \bibinfo
  {author} {\bibfnamefont {M.}~\bibnamefont {{Ashdown}}}, \bibinfo {author}
  {\bibfnamefont {F.}~\bibnamefont {{Atrio-Barandela}}}, \bibinfo {author}
  {\bibfnamefont {J.}~\bibnamefont {{Aumont}}}, \bibinfo {author}
  {\bibfnamefont {C.}~\bibnamefont {{Baccigalupi}}}, \bibinfo {author}
  {\bibfnamefont {A.~J.}\ \bibnamefont {{Banday}}}, \ and\ \bibinfo {author}
  {\bibnamefont {et~al.}},\ }\href@noop {} {\bibfield  {journal} {\bibinfo
  {journal} {ArXiv e-prints}\ } (\bibinfo {year} {2013}{\natexlab{a}})},\
  \Eprint {http://arxiv.org/abs/1303.5082} {arXiv:1303.5082 [astro-ph.CO]}
  \BibitemShut {NoStop}%
\bibitem [{\citenamefont {{Planck Collaboration}}\ \emph
  {et~al.}(2013{\natexlab{b}})\citenamefont {{Planck Collaboration}},
  \citenamefont {{Ade}}, \citenamefont {{Aghanim}}, \citenamefont
  {{Armitage-Caplan}}, \citenamefont {{Arnaud}}, \citenamefont {{Ashdown}},
  \citenamefont {{Atrio-Barandela}}, \citenamefont {{Aumont}}, \citenamefont
  {{Baccigalupi}}, \citenamefont {{Banday}},\ and\ \citenamefont
  {et~al.}}]{2013arXiv1303.5084P}%
  \BibitemOpen
  \bibfield  {author} {\bibinfo {author} {\bibnamefont {{Planck
  Collaboration}}}, \bibinfo {author} {\bibfnamefont {P.~A.~R.}\ \bibnamefont
  {{Ade}}}, \bibinfo {author} {\bibfnamefont {N.}~\bibnamefont {{Aghanim}}},
  \bibinfo {author} {\bibfnamefont {C.}~\bibnamefont {{Armitage-Caplan}}},
  \bibinfo {author} {\bibfnamefont {M.}~\bibnamefont {{Arnaud}}}, \bibinfo
  {author} {\bibfnamefont {M.}~\bibnamefont {{Ashdown}}}, \bibinfo {author}
  {\bibfnamefont {F.}~\bibnamefont {{Atrio-Barandela}}}, \bibinfo {author}
  {\bibfnamefont {J.}~\bibnamefont {{Aumont}}}, \bibinfo {author}
  {\bibfnamefont {C.}~\bibnamefont {{Baccigalupi}}}, \bibinfo {author}
  {\bibfnamefont {A.~J.}\ \bibnamefont {{Banday}}}, \ and\ \bibinfo {author}
  {\bibnamefont {et~al.}},\ }\href@noop {} {\bibfield  {journal} {\bibinfo
  {journal} {ArXiv e-prints}\ } (\bibinfo {year} {2013}{\natexlab{b}})},\
  \Eprint {http://arxiv.org/abs/1303.5084} {arXiv:1303.5084 [astro-ph.CO]}
  \BibitemShut {NoStop}%
\bibitem [{\citenamefont {{McAllister}}\ and\ \citenamefont
  {{Silverstein}}(2008)}]{2008GReGr..40..565M}%
  \BibitemOpen
  \bibfield  {author} {\bibinfo {author} {\bibfnamefont {L.}~\bibnamefont
  {{McAllister}}}\ and\ \bibinfo {author} {\bibfnamefont {E.}~\bibnamefont
  {{Silverstein}}},\ }\href {\doibase 10.1007/s10714-007-0556-6} {\bibfield
  {journal} {\bibinfo  {journal} {General Relativity and Gravitation}\ }\textbf
  {\bibinfo {volume} {40}},\ \bibinfo {pages} {565} (\bibinfo {year} {2008})},\
  \Eprint {http://arxiv.org/abs/0710.2951} {arXiv:0710.2951 [hep-th]}
  \BibitemShut {NoStop}%
\bibitem [{\citenamefont {{Behbahani}}\ \emph {et~al.}(2012)\citenamefont
  {{Behbahani}}, \citenamefont {{Dymarsky}}, \citenamefont {{Mirbabayi}},\ and\
  \citenamefont {{Senatore}}}]{2012JCAP...12..036B}%
  \BibitemOpen
  \bibfield  {author} {\bibinfo {author} {\bibfnamefont {S.~R.}\ \bibnamefont
  {{Behbahani}}}, \bibinfo {author} {\bibfnamefont {A.}~\bibnamefont
  {{Dymarsky}}}, \bibinfo {author} {\bibfnamefont {M.}~\bibnamefont
  {{Mirbabayi}}}, \ and\ \bibinfo {author} {\bibfnamefont {L.}~\bibnamefont
  {{Senatore}}},\ }\href {\doibase 10.1088/1475-7516/2012/12/036} {\bibfield
  {journal} {\bibinfo  {journal} {\jcap}\ }\textbf {\bibinfo {volume} {12}},\
  \bibinfo {eid} {036} (\bibinfo {year} {2012})},\ \Eprint
  {http://arxiv.org/abs/1111.3373} {arXiv:1111.3373 [hep-th]} \BibitemShut
  {NoStop}%
\bibitem [{\citenamefont {Freese}\ \emph {et~al.}(1990)\citenamefont {Freese},
  \citenamefont {Frieman},\ and\ \citenamefont {Olinto}}]{PhysRevLett.65.3233}%
  \BibitemOpen
  \bibfield  {author} {\bibinfo {author} {\bibfnamefont {K.}~\bibnamefont
  {Freese}}, \bibinfo {author} {\bibfnamefont {J.~A.}\ \bibnamefont {Frieman}},
  \ and\ \bibinfo {author} {\bibfnamefont {A.~V.}\ \bibnamefont {Olinto}},\
  }\href {\doibase 10.1103/PhysRevLett.65.3233} {\bibfield  {journal} {\bibinfo
   {journal} {Phys. Rev. Lett.}\ }\textbf {\bibinfo {volume} {65}},\ \bibinfo
  {pages} {3233} (\bibinfo {year} {1990})}\BibitemShut {NoStop}%
\bibitem [{\citenamefont {{Silverstein}}\ and\ \citenamefont
  {{Westphal}}(2008)}]{2008PhRvD..78j6003S}%
  \BibitemOpen
  \bibfield  {author} {\bibinfo {author} {\bibfnamefont {E.}~\bibnamefont
  {{Silverstein}}}\ and\ \bibinfo {author} {\bibfnamefont {A.}~\bibnamefont
  {{Westphal}}},\ }\href {\doibase 10.1103/PhysRevD.78.106003} {\bibfield
  {journal} {\bibinfo  {journal} {\prd}\ }\textbf {\bibinfo {volume} {78}},\
  \bibinfo {eid} {106003} (\bibinfo {year} {2008})},\ \Eprint
  {http://arxiv.org/abs/0803.3085} {arXiv:0803.3085 [hep-th]} \BibitemShut
  {NoStop}%
\bibitem [{\citenamefont {{Flauger}}\ \emph {et~al.}(2010)\citenamefont
  {{Flauger}}, \citenamefont {{McAllister}}, \citenamefont {{Pajer}},
  \citenamefont {{Westphal}},\ and\ \citenamefont
  {{Xu}}}]{2010JCAP...06..009F}%
  \BibitemOpen
  \bibfield  {author} {\bibinfo {author} {\bibfnamefont {R.}~\bibnamefont
  {{Flauger}}}, \bibinfo {author} {\bibfnamefont {L.}~\bibnamefont
  {{McAllister}}}, \bibinfo {author} {\bibfnamefont {E.}~\bibnamefont
  {{Pajer}}}, \bibinfo {author} {\bibfnamefont {A.}~\bibnamefont {{Westphal}}},
  \ and\ \bibinfo {author} {\bibfnamefont {G.}~\bibnamefont {{Xu}}},\ }\href
  {\doibase 10.1088/1475-7516/2010/06/009} {\bibfield  {journal} {\bibinfo
  {journal} {\jcap}\ }\textbf {\bibinfo {volume} {6}},\ \bibinfo {eid} {009}
  (\bibinfo {year} {2010})},\ \Eprint {http://arxiv.org/abs/0907.2916}
  {arXiv:0907.2916 [hep-th]} \BibitemShut {NoStop}%
\bibitem [{\citenamefont {{Flauger}}\ and\ \citenamefont
  {{Pajer}}(2011)}]{2011JCAP...01..017F}%
  \BibitemOpen
  \bibfield  {author} {\bibinfo {author} {\bibfnamefont {R.}~\bibnamefont
  {{Flauger}}}\ and\ \bibinfo {author} {\bibfnamefont {E.}~\bibnamefont
  {{Pajer}}},\ }\href {\doibase 10.1088/1475-7516/2011/01/017} {\bibfield
  {journal} {\bibinfo  {journal} {\jcap}\ }\textbf {\bibinfo {volume} {1}},\
  \bibinfo {eid} {017} (\bibinfo {year} {2011})},\ \Eprint
  {http://arxiv.org/abs/1002.0833} {arXiv:1002.0833 [hep-th]} \BibitemShut
  {NoStop}%
\bibitem [{\citenamefont {{Greene}}\ \emph {et~al.}(2005)\citenamefont
  {{Greene}}, \citenamefont {{Schalm}}, \citenamefont {{van der Schaar}},\ and\
  \citenamefont {{Shiu}}}]{2005tsra.conf....1G}%
  \BibitemOpen
  \bibfield  {author} {\bibinfo {author} {\bibfnamefont {B.}~\bibnamefont
  {{Greene}}}, \bibinfo {author} {\bibfnamefont {K.}~\bibnamefont {{Schalm}}},
  \bibinfo {author} {\bibfnamefont {J.~P.}\ \bibnamefont {{van der Schaar}}}, \
  and\ \bibinfo {author} {\bibfnamefont {G.}~\bibnamefont {{Shiu}}},\ }in\
  \href@noop {} {\emph {\bibinfo {booktitle} {22nd Texas Symposium on
  Relativistic Astrophysics}}},\ \bibinfo {editor} {edited by\ \bibinfo
  {editor} {\bibfnamefont {P.}~\bibnamefont {{Chen}}}, \bibinfo {editor}
  {\bibfnamefont {E.}~\bibnamefont {{Bloom}}}, \bibinfo {editor} {\bibfnamefont
  {G.}~\bibnamefont {{Madejski}}}, \ and\ \bibinfo {editor} {\bibfnamefont
  {V.}~\bibnamefont {{Patrosian}}}}\ (\bibinfo {year} {2005})\ pp.\ \bibinfo
  {pages} {1--8},\ \Eprint {http://arxiv.org/abs/arXiv:astro-ph/0503458}
  {arXiv:astro-ph/0503458} \BibitemShut {NoStop}%
\bibitem [{\citenamefont {{Meerburg}}\ \emph {et~al.}(2009)\citenamefont
  {{Meerburg}}, \citenamefont {{van der Schaar}},\ and\ \citenamefont
  {{Corasaniti}}}]{2009JCAP...05..018M}%
  \BibitemOpen
  \bibfield  {author} {\bibinfo {author} {\bibfnamefont {P.~D.}\ \bibnamefont
  {{Meerburg}}}, \bibinfo {author} {\bibfnamefont {J.~P.}\ \bibnamefont {{van
  der Schaar}}}, \ and\ \bibinfo {author} {\bibfnamefont {P.~S.}\ \bibnamefont
  {{Corasaniti}}},\ }\href {\doibase 10.1088/1475-7516/2009/05/018} {\bibfield
  {journal} {\bibinfo  {journal} {\jcap}\ }\textbf {\bibinfo {volume} {5}},\
  \bibinfo {eid} {018} (\bibinfo {year} {2009})},\ \Eprint
  {http://arxiv.org/abs/0901.4044} {arXiv:0901.4044 [hep-th]} \BibitemShut
  {NoStop}%
\bibitem [{\citenamefont {{D'Amico}}\ \emph {et~al.}(2013)\citenamefont
  {{D'Amico}}, \citenamefont {{Gobbetti}}, \citenamefont {{Kleban}},\ and\
  \citenamefont {{Schillo}}}]{2013JCAP...03..004D}%
  \BibitemOpen
  \bibfield  {author} {\bibinfo {author} {\bibfnamefont {G.}~\bibnamefont
  {{D'Amico}}}, \bibinfo {author} {\bibfnamefont {R.}~\bibnamefont
  {{Gobbetti}}}, \bibinfo {author} {\bibfnamefont {M.}~\bibnamefont
  {{Kleban}}}, \ and\ \bibinfo {author} {\bibfnamefont {M.}~\bibnamefont
  {{Schillo}}},\ }\href {\doibase 10.1088/1475-7516/2013/03/004} {\bibfield
  {journal} {\bibinfo  {journal} {\jcap}\ }\textbf {\bibinfo {volume} {3}},\
  \bibinfo {eid} {004} (\bibinfo {year} {2013})},\ \Eprint
  {http://arxiv.org/abs/1211.4589} {arXiv:1211.4589 [hep-th]} \BibitemShut
  {NoStop}%
\bibitem [{\citenamefont {{Ach{\'u}carro}}\ \emph {et~al.}(2011)\citenamefont
  {{Ach{\'u}carro}}, \citenamefont {{Gong}}, \citenamefont {{Hardeman}},
  \citenamefont {{Palma}},\ and\ \citenamefont
  {{Patil}}}]{2011JCAP...01..030A}%
  \BibitemOpen
  \bibfield  {author} {\bibinfo {author} {\bibfnamefont {A.}~\bibnamefont
  {{Ach{\'u}carro}}}, \bibinfo {author} {\bibfnamefont {J.-O.}\ \bibnamefont
  {{Gong}}}, \bibinfo {author} {\bibfnamefont {S.}~\bibnamefont {{Hardeman}}},
  \bibinfo {author} {\bibfnamefont {G.~A.}\ \bibnamefont {{Palma}}}, \ and\
  \bibinfo {author} {\bibfnamefont {S.~P.}\ \bibnamefont {{Patil}}},\ }\href
  {\doibase 10.1088/1475-7516/2011/01/030} {\bibfield  {journal} {\bibinfo
  {journal} {\jcap}\ }\textbf {\bibinfo {volume} {1}},\ \bibinfo {eid} {030}
  (\bibinfo {year} {2011})},\ \Eprint {http://arxiv.org/abs/1010.3693}
  {arXiv:1010.3693 [hep-ph]} \BibitemShut {NoStop}%
\bibitem [{\citenamefont {{Battefeld}}\ \emph {et~al.}(2013)\citenamefont
  {{Battefeld}}, \citenamefont {{Niemeyer}},\ and\ \citenamefont
  {{Vlaykov}}}]{2013JCAP...05..006B}%
  \BibitemOpen
  \bibfield  {author} {\bibinfo {author} {\bibfnamefont {T.}~\bibnamefont
  {{Battefeld}}}, \bibinfo {author} {\bibfnamefont {J.~C.}\ \bibnamefont
  {{Niemeyer}}}, \ and\ \bibinfo {author} {\bibfnamefont {D.}~\bibnamefont
  {{Vlaykov}}},\ }\href {\doibase 10.1088/1475-7516/2013/05/006} {\bibfield
  {journal} {\bibinfo  {journal} {\jcap}\ }\textbf {\bibinfo {volume} {5}},\
  \bibinfo {eid} {006} (\bibinfo {year} {2013})},\ \Eprint
  {http://arxiv.org/abs/1302.3877} {arXiv:1302.3877 [astro-ph.CO]} \BibitemShut
  {NoStop}%
\bibitem [{\citenamefont {{Martin}}\ and\ \citenamefont
  {{Ringeval}}(2004)}]{2004PhRvD..69h3515M}%
  \BibitemOpen
  \bibfield  {author} {\bibinfo {author} {\bibfnamefont {J.}~\bibnamefont
  {{Martin}}}\ and\ \bibinfo {author} {\bibfnamefont {C.}~\bibnamefont
  {{Ringeval}}},\ }\href {\doibase 10.1103/PhysRevD.69.083515} {\bibfield
  {journal} {\bibinfo  {journal} {\prd}\ }\textbf {\bibinfo {volume} {69}},\
  \bibinfo {eid} {083515} (\bibinfo {year} {2004})},\ \Eprint
  {http://arxiv.org/abs/arXiv:astro-ph/0310382} {arXiv:astro-ph/0310382}
  \BibitemShut {NoStop}%
\bibitem [{\citenamefont {{Hamann}}\ \emph {et~al.}(2007)\citenamefont
  {{Hamann}}, \citenamefont {{Covi}}, \citenamefont {{Melchiorri}},\ and\
  \citenamefont {{Slosar}}}]{2007PhRvD..76b3503H}%
  \BibitemOpen
  \bibfield  {author} {\bibinfo {author} {\bibfnamefont {J.}~\bibnamefont
  {{Hamann}}}, \bibinfo {author} {\bibfnamefont {L.}~\bibnamefont {{Covi}}},
  \bibinfo {author} {\bibfnamefont {A.}~\bibnamefont {{Melchiorri}}}, \ and\
  \bibinfo {author} {\bibfnamefont {A.}~\bibnamefont {{Slosar}}},\ }\href
  {\doibase 10.1103/PhysRevD.76.023503} {\bibfield  {journal} {\bibinfo
  {journal} {\prd}\ }\textbf {\bibinfo {volume} {76}},\ \bibinfo {eid} {023503}
  (\bibinfo {year} {2007})},\ \Eprint
  {http://arxiv.org/abs/arXiv:astro-ph/0701380} {arXiv:astro-ph/0701380}
  \BibitemShut {NoStop}%
\bibitem [{\citenamefont {{Hamann}}\ \emph {et~al.}(2010)\citenamefont
  {{Hamann}}, \citenamefont {{Shafieloo}},\ and\ \citenamefont
  {{Souradeep}}}]{2010JCAP...04..010H}%
  \BibitemOpen
  \bibfield  {author} {\bibinfo {author} {\bibfnamefont {J.}~\bibnamefont
  {{Hamann}}}, \bibinfo {author} {\bibfnamefont {A.}~\bibnamefont
  {{Shafieloo}}}, \ and\ \bibinfo {author} {\bibfnamefont {T.}~\bibnamefont
  {{Souradeep}}},\ }\href {\doibase 10.1088/1475-7516/2010/04/010} {\bibfield
  {journal} {\bibinfo  {journal} {\jcap}\ }\textbf {\bibinfo {volume} {4}},\
  \bibinfo {eid} {010} (\bibinfo {year} {2010})},\ \Eprint
  {http://arxiv.org/abs/0912.2728} {arXiv:0912.2728 [astro-ph.CO]} \BibitemShut
  {NoStop}%
\bibitem [{\citenamefont {{Dvorkin}}\ and\ \citenamefont
  {{Hu}}(2011)}]{2011PhRvD..84f3515D}%
  \BibitemOpen
  \bibfield  {author} {\bibinfo {author} {\bibfnamefont {C.}~\bibnamefont
  {{Dvorkin}}}\ and\ \bibinfo {author} {\bibfnamefont {W.}~\bibnamefont
  {{Hu}}},\ }\href {\doibase 10.1103/PhysRevD.84.063515} {\bibfield  {journal}
  {\bibinfo  {journal} {\prd}\ }\textbf {\bibinfo {volume} {84}},\ \bibinfo
  {eid} {063515} (\bibinfo {year} {2011})},\ \Eprint
  {http://arxiv.org/abs/1106.4016} {arXiv:1106.4016 [astro-ph.CO]} \BibitemShut
  {NoStop}%
\bibitem [{\citenamefont {{Meerburg}}\ \emph {et~al.}(2012)\citenamefont
  {{Meerburg}}, \citenamefont {{Wijers}},\ and\ \citenamefont {{van der
  Schaar}}}]{2012MNRAS.421..369M}%
  \BibitemOpen
  \bibfield  {author} {\bibinfo {author} {\bibfnamefont {P.~D.}\ \bibnamefont
  {{Meerburg}}}, \bibinfo {author} {\bibfnamefont {R.~A.~M.~J.}\ \bibnamefont
  {{Wijers}}}, \ and\ \bibinfo {author} {\bibfnamefont {J.~P.}\ \bibnamefont
  {{van der Schaar}}},\ }\href {\doibase 10.1111/j.1365-2966.2011.20311.x}
  {\bibfield  {journal} {\bibinfo  {journal} {\mnras}\ }\textbf {\bibinfo
  {volume} {421}},\ \bibinfo {pages} {369} (\bibinfo {year} {2012})},\ \Eprint
  {http://arxiv.org/abs/1109.5264} {arXiv:1109.5264 [astro-ph.CO]} \BibitemShut
  {NoStop}%
\bibitem [{\citenamefont {{Benetti}}\ \emph {et~al.}(2011)\citenamefont
  {{Benetti}}, \citenamefont {{Lattanzi}}, \citenamefont {{Calabrese}},\ and\
  \citenamefont {{Melchiorri}}}]{2011PhRvD..84f3509B}%
  \BibitemOpen
  \bibfield  {author} {\bibinfo {author} {\bibfnamefont {M.}~\bibnamefont
  {{Benetti}}}, \bibinfo {author} {\bibfnamefont {M.}~\bibnamefont
  {{Lattanzi}}}, \bibinfo {author} {\bibfnamefont {E.}~\bibnamefont
  {{Calabrese}}}, \ and\ \bibinfo {author} {\bibfnamefont {A.}~\bibnamefont
  {{Melchiorri}}},\ }\href {\doibase 10.1103/PhysRevD.84.063509} {\bibfield
  {journal} {\bibinfo  {journal} {\prd}\ }\textbf {\bibinfo {volume} {84}},\
  \bibinfo {eid} {063509} (\bibinfo {year} {2011})},\ \Eprint
  {http://arxiv.org/abs/1107.4992} {arXiv:1107.4992 [astro-ph.CO]} \BibitemShut
  {NoStop}%
\bibitem [{\citenamefont {{Aich}}\ \emph {et~al.}(2013)\citenamefont {{Aich}},
  \citenamefont {{Hazra}}, \citenamefont {{Sriramkumar}},\ and\ \citenamefont
  {{Souradeep}}}]{2013PhRvD..87h3526A}%
  \BibitemOpen
  \bibfield  {author} {\bibinfo {author} {\bibfnamefont {M.}~\bibnamefont
  {{Aich}}}, \bibinfo {author} {\bibfnamefont {D.~K.}\ \bibnamefont {{Hazra}}},
  \bibinfo {author} {\bibfnamefont {L.}~\bibnamefont {{Sriramkumar}}}, \ and\
  \bibinfo {author} {\bibfnamefont {T.}~\bibnamefont {{Souradeep}}},\ }\href
  {\doibase 10.1103/PhysRevD.87.083526} {\bibfield  {journal} {\bibinfo
  {journal} {\prd}\ }\textbf {\bibinfo {volume} {87}},\ \bibinfo {eid} {083526}
  (\bibinfo {year} {2013})}\BibitemShut {NoStop}%
\bibitem [{\citenamefont {{Peiris}}\ \emph {et~al.}(2013)\citenamefont
  {{Peiris}}, \citenamefont {{Easther}},\ and\ \citenamefont
  {{Flauger}}}]{2013arXiv1303.2616P}%
  \BibitemOpen
  \bibfield  {author} {\bibinfo {author} {\bibfnamefont {H.}~\bibnamefont
  {{Peiris}}}, \bibinfo {author} {\bibfnamefont {R.}~\bibnamefont {{Easther}}},
  \ and\ \bibinfo {author} {\bibfnamefont {R.}~\bibnamefont {{Flauger}}},\
  }\href@noop {} {\bibfield  {journal} {\bibinfo  {journal} {ArXiv e-prints}\ }
  (\bibinfo {year} {2013})},\ \Eprint {http://arxiv.org/abs/1303.2616}
  {arXiv:1303.2616 [astro-ph.CO]} \BibitemShut {NoStop}%
\bibitem [{\citenamefont {{Bennett}}\ \emph {et~al.}(2012)\citenamefont
  {{Bennett}}, \citenamefont {{Larson}}, \citenamefont {{Weiland}},
  \citenamefont {{Jarosik}}, \citenamefont {{Hinshaw}}, \citenamefont
  {{Odegard}}, \citenamefont {{Smith}}, \citenamefont {{Hill}}, \citenamefont
  {{Gold}}, \citenamefont {{Halpern}}, \citenamefont {{Komatsu}}, \citenamefont
  {{Nolta}}, \citenamefont {{Page}}, \citenamefont {{Spergel}}, \citenamefont
  {{Wollack}}, \citenamefont {{Dunkley}}, \citenamefont {{Kogut}},
  \citenamefont {{Limon}}, \citenamefont {{Meyer}}, \citenamefont {{Tucker}},\
  and\ \citenamefont {{Wright}}}]{2012arXiv1212.5225B}%
  \BibitemOpen
  \bibfield  {author} {\bibinfo {author} {\bibfnamefont {C.~L.}\ \bibnamefont
  {{Bennett}}}, \bibinfo {author} {\bibfnamefont {D.}~\bibnamefont {{Larson}}},
  \bibinfo {author} {\bibfnamefont {J.~L.}\ \bibnamefont {{Weiland}}}, \bibinfo
  {author} {\bibfnamefont {N.}~\bibnamefont {{Jarosik}}}, \bibinfo {author}
  {\bibfnamefont {G.}~\bibnamefont {{Hinshaw}}}, \bibinfo {author}
  {\bibfnamefont {N.}~\bibnamefont {{Odegard}}}, \bibinfo {author}
  {\bibfnamefont {K.~M.}\ \bibnamefont {{Smith}}}, \bibinfo {author}
  {\bibfnamefont {R.~S.}\ \bibnamefont {{Hill}}}, \bibinfo {author}
  {\bibfnamefont {B.}~\bibnamefont {{Gold}}}, \bibinfo {author} {\bibfnamefont
  {M.}~\bibnamefont {{Halpern}}}, \bibinfo {author} {\bibfnamefont
  {E.}~\bibnamefont {{Komatsu}}}, \bibinfo {author} {\bibfnamefont {M.~R.}\
  \bibnamefont {{Nolta}}}, \bibinfo {author} {\bibfnamefont {L.}~\bibnamefont
  {{Page}}}, \bibinfo {author} {\bibfnamefont {D.~N.}\ \bibnamefont
  {{Spergel}}}, \bibinfo {author} {\bibfnamefont {E.}~\bibnamefont
  {{Wollack}}}, \bibinfo {author} {\bibfnamefont {J.}~\bibnamefont
  {{Dunkley}}}, \bibinfo {author} {\bibfnamefont {A.}~\bibnamefont {{Kogut}}},
  \bibinfo {author} {\bibfnamefont {M.}~\bibnamefont {{Limon}}}, \bibinfo
  {author} {\bibfnamefont {S.~S.}\ \bibnamefont {{Meyer}}}, \bibinfo {author}
  {\bibfnamefont {G.~S.}\ \bibnamefont {{Tucker}}}, \ and\ \bibinfo {author}
  {\bibfnamefont {E.~L.}\ \bibnamefont {{Wright}}},\ }\href@noop {} {\bibfield
  {journal} {\bibinfo  {journal} {ArXiv e-prints}\ } (\bibinfo {year}
  {2012})},\ \Eprint {http://arxiv.org/abs/1212.5225} {arXiv:1212.5225
  [astro-ph.CO]} \BibitemShut {NoStop}%
\bibitem [{\citenamefont {{Chen}}\ \emph {et~al.}(2007)\citenamefont {{Chen}},
  \citenamefont {{Easther}},\ and\ \citenamefont
  {{Lim}}}]{2007JCAP...06..023C}%
  \BibitemOpen
  \bibfield  {author} {\bibinfo {author} {\bibfnamefont {X.}~\bibnamefont
  {{Chen}}}, \bibinfo {author} {\bibfnamefont {R.}~\bibnamefont {{Easther}}}, \
  and\ \bibinfo {author} {\bibfnamefont {E.~A.}\ \bibnamefont {{Lim}}},\ }\href
  {\doibase 10.1088/1475-7516/2007/06/023} {\bibfield  {journal} {\bibinfo
  {journal} {\jcap}\ }\textbf {\bibinfo {volume} {6}},\ \bibinfo {eid} {023}
  (\bibinfo {year} {2007})},\ \Eprint
  {http://arxiv.org/abs/arXiv:astro-ph/0611645} {arXiv:astro-ph/0611645}
  \BibitemShut {NoStop}%
\bibitem [{\citenamefont {{Planck Collaboration}}\ \emph
  {et~al.}(2013{\natexlab{c}})\citenamefont {{Planck Collaboration}},
  \citenamefont {{Ade}}, \citenamefont {{Aghanim}}, \citenamefont
  {{Armitage-Caplan}}, \citenamefont {{Arnaud}}, \citenamefont {{Ashdown}},
  \citenamefont {{Atrio-Barandela}}, \citenamefont {{Aumont}}, \citenamefont
  {{Baccigalupi}}, \citenamefont {{Banday}},\ and\ \citenamefont
  {et~al.}}]{2013arXiv1303.5076P}%
  \BibitemOpen
  \bibfield  {author} {\bibinfo {author} {\bibnamefont {{Planck
  Collaboration}}}, \bibinfo {author} {\bibfnamefont {P.~A.~R.}\ \bibnamefont
  {{Ade}}}, \bibinfo {author} {\bibfnamefont {N.}~\bibnamefont {{Aghanim}}},
  \bibinfo {author} {\bibfnamefont {C.}~\bibnamefont {{Armitage-Caplan}}},
  \bibinfo {author} {\bibfnamefont {M.}~\bibnamefont {{Arnaud}}}, \bibinfo
  {author} {\bibfnamefont {M.}~\bibnamefont {{Ashdown}}}, \bibinfo {author}
  {\bibfnamefont {F.}~\bibnamefont {{Atrio-Barandela}}}, \bibinfo {author}
  {\bibfnamefont {J.}~\bibnamefont {{Aumont}}}, \bibinfo {author}
  {\bibfnamefont {C.}~\bibnamefont {{Baccigalupi}}}, \bibinfo {author}
  {\bibfnamefont {A.~J.}\ \bibnamefont {{Banday}}}, \ and\ \bibinfo {author}
  {\bibnamefont {et~al.}},\ }\href@noop {} {\bibfield  {journal} {\bibinfo
  {journal} {ArXiv e-prints}\ } (\bibinfo {year} {2013}{\natexlab{c}})},\
  \Eprint {http://arxiv.org/abs/1303.5076} {arXiv:1303.5076 [astro-ph.CO]}
  \BibitemShut {NoStop}%
\bibitem [{\citenamefont {{Feroz}}\ \emph {et~al.}(2009)\citenamefont
  {{Feroz}}, \citenamefont {{Hobson}},\ and\ \citenamefont
  {{Bridges}}}]{2009MNRAS.398.1601F}%
  \BibitemOpen
  \bibfield  {author} {\bibinfo {author} {\bibfnamefont {F.}~\bibnamefont
  {{Feroz}}}, \bibinfo {author} {\bibfnamefont {M.~P.}\ \bibnamefont
  {{Hobson}}}, \ and\ \bibinfo {author} {\bibfnamefont {M.}~\bibnamefont
  {{Bridges}}},\ }\href {\doibase 10.1111/j.1365-2966.2009.14548.x} {\bibfield
  {journal} {\bibinfo  {journal} {\mnras}\ }\textbf {\bibinfo {volume} {398}},\
  \bibinfo {pages} {1601} (\bibinfo {year} {2009})},\ \Eprint
  {http://arxiv.org/abs/0809.3437} {arXiv:0809.3437} \BibitemShut {NoStop}%
\bibitem [{\citenamefont {Meerburg}\ \emph {et~al.}(2011)\citenamefont
  {Meerburg}, \citenamefont {Wijers},\ and\ \citenamefont {van~der
  Schaar}}]{Meerburg:2011gd}%
  \BibitemOpen
  \bibfield  {author} {\bibinfo {author} {\bibfnamefont {P.~D.}\ \bibnamefont
  {Meerburg}}, \bibinfo {author} {\bibfnamefont {R.}~\bibnamefont {Wijers}}, \
  and\ \bibinfo {author} {\bibfnamefont {J.~P.}\ \bibnamefont {van~der
  Schaar}},\ }\href@noop {} {\  (\bibinfo {year} {2011})},\ \Eprint
  {http://arxiv.org/abs/1109.5264} {arXiv:1109.5264 [astro-ph.CO]} \BibitemShut
  {NoStop}%
\bibitem [{\citenamefont {{Lewis}}\ and\ \citenamefont
  {{Bridle}}(2002)}]{2002PhRvD..66j3511L}%
  \BibitemOpen
  \bibfield  {author} {\bibinfo {author} {\bibfnamefont {A.}~\bibnamefont
  {{Lewis}}}\ and\ \bibinfo {author} {\bibfnamefont {S.}~\bibnamefont
  {{Bridle}}},\ }\href {\doibase 10.1103/PhysRevD.66.103511} {\bibfield
  {journal} {\bibinfo  {journal} {\prd}\ }\textbf {\bibinfo {volume} {66}},\
  \bibinfo {eid} {103511} (\bibinfo {year} {2002})},\ \Eprint
  {http://arxiv.org/abs/arXiv:astro-ph/0205436} {arXiv:astro-ph/0205436}
  \BibitemShut {NoStop}%
\bibitem [{\citenamefont {{Perotto}}\ \emph {et~al.}(2006)\citenamefont
  {{Perotto}}, \citenamefont {{Lesgourgues}}, \citenamefont {{Hannestad}},
  \citenamefont {{Tu}},\ and\ \citenamefont {{Y Y
  Wong}}}]{2006JCAP...10..013P}%
  \BibitemOpen
  \bibfield  {author} {\bibinfo {author} {\bibfnamefont {L.}~\bibnamefont
  {{Perotto}}}, \bibinfo {author} {\bibfnamefont {J.}~\bibnamefont
  {{Lesgourgues}}}, \bibinfo {author} {\bibfnamefont {S.}~\bibnamefont
  {{Hannestad}}}, \bibinfo {author} {\bibfnamefont {H.}~\bibnamefont {{Tu}}}, \
  and\ \bibinfo {author} {\bibfnamefont {Y.}~\bibnamefont {{Y Y Wong}}},\
  }\href {\doibase 10.1088/1475-7516/2006/10/013} {\bibfield  {journal}
  {\bibinfo  {journal} {\jcap}\ }\textbf {\bibinfo {volume} {10}},\ \bibinfo
  {eid} {013} (\bibinfo {year} {2006})},\ \Eprint
  {http://arxiv.org/abs/arXiv:astro-ph/0606227} {arXiv:astro-ph/0606227}
  \BibitemShut {NoStop}%
\bibitem [{\citenamefont {{Liddle}}(2007)}]{2007MNRAS.377L..74L}%
  \BibitemOpen
  \bibfield  {author} {\bibinfo {author} {\bibfnamefont {A.~R.}\ \bibnamefont
  {{Liddle}}},\ }\href {\doibase 10.1111/j.1745-3933.2007.00306.x} {\bibfield
  {journal} {\bibinfo  {journal} {\mnras}\ }\textbf {\bibinfo {volume} {377}},\
  \bibinfo {pages} {L74} (\bibinfo {year} {2007})},\ \Eprint
  {http://arxiv.org/abs/arXiv:astro-ph/0701113} {arXiv:astro-ph/0701113}
  \BibitemShut {NoStop}%
\bibitem [{\citenamefont {{Melia}}\ and\ \citenamefont
  {{Maier}}(2013)}]{2013MNRAS.tmp.1367M}%
  \BibitemOpen
  \bibfield  {author} {\bibinfo {author} {\bibfnamefont {F.}~\bibnamefont
  {{Melia}}}\ and\ \bibinfo {author} {\bibfnamefont {R.~S.}\ \bibnamefont
  {{Maier}}},\ }\href {\doibase 10.1093/mnras/stt596} {\bibfield  {journal}
  {\bibinfo  {journal} {\mnras}\ } (\bibinfo {year} {2013}),\
  10.1093/mnras/stt596},\ \Eprint {http://arxiv.org/abs/1304.1802}
  {arXiv:1304.1802 [astro-ph.CO]} \BibitemShut {NoStop}%
\bibitem [{\citenamefont {{Easther}}\ and\ \citenamefont
  {{Flauger}}(2013)}]{2013Flaugera}%
  \BibitemOpen
  \bibfield  {author} {\bibinfo {author} {\bibfnamefont {R.}~\bibnamefont
  {{Easther}}}\ and\ \bibinfo {author} {\bibfnamefont {R.}~\bibnamefont
  {{Flauger}}},\ }\href@noop {} {\bibfield  {journal} {\bibinfo  {journal}
  {ArXiv e-prints}\ } (\bibinfo {year} {2013})},\ \Eprint
  {http://arxiv.org/abs/1308.3736} {arXiv:1308.3736 [astro-ph.CO]} \BibitemShut
  {NoStop}%
\end{thebibliography}%

\end{document}